\documentclass[a4paper,fleqn]{cas-sc}

\usepackage{subfigure}
\usepackage{natbib}
\setcitestyle{numbers,square}

\def\tsc#1{\csdef{#1}{\textsc{\lowercase{#1}}\xspace}}
\tsc{WGM}
\tsc{QE}
\tsc{EP}
\tsc{PMS}
\tsc{BEC}
\tsc{DE}

\begin{document}
\let\WriteBookmarks\relax
\def\floatpagepagefraction{1}
\def\textpagefraction{.001}

\shorttitle{Adaptive Payoff-driven Interaction in Networked Snowdrift Games}

\shortauthors{Xiaojin Xiong et~al.}

\title [mode = title]{Adaptive Payoff-driven Interaction in Networked Snowdrift Games}
%

\author[1]{Xiaojin Xiong}[style=Chinese]

\author[1]{Yichao Yao}[style=chinese]

\author[1]{Minyu Feng}[style=chinese,orcid=0000-0001-6772-3017]
\cormark[1]

\ead{myfeng@swu.edu.cn}

\author[2,3]{Manuel Chica}[orcid=0000-0002-4717-1056]

\affiliation[1]{organization={College of Artificial Intelligence},
    addressline={Southwest University}, 
    city={Chongqing},
    postcode={400715}, 
    country={PR China}}
\affiliation[2]{organization={Andalusian Research Institute DaSCI “Data Science and Computational Intelligence”},
    addressline={University of Granada}, 
    city={Granada},
    postcode={18071}, 
    country={Spain}}
\affiliation[3]{organization={School of Information and Physical Sciences},
    addressline={The University of Newcastle}, 
    city={Callaghan},
    postcode={NSW 2308}, 
    country={Australia}}

\begin{abstract}
In social dilemmas, most interactions are transient and susceptible to restructuring, leading to continuous changes in social networks over time. Typically, agents assess the rewards of their current interactions and adjust their connections to optimize outcomes. In this paper, we introduce an adaptive network model in the snowdrift game to examine dynamic levels of cooperation and network topology, involving the potential for both the termination of existing connections and the establishment of new ones. In particular, we define the agent's asymmetric disassociation tendency toward their neighbors, which fundamentally determines the probability of edge dismantlement. The mechanism allows agents to selectively sever and rewire their connections to alternative individuals to refine partnerships. Our findings reveal that adaptive networks are particularly effective in promoting a robust evolution toward states of either pure cooperation or complete defection, especially under conditions of extreme cost-benefit ratios, as compared to static network models. Moreover, the dynamic restructuring of connections and the distribution of network degrees among agents are closely linked to the levels of cooperation in stationary states. Specifically, cooperators tend to seek broader neighborhoods when confronted with the invasion of multiple defectors.
\end{abstract}



\maketitle

\section{Introduction}
The exploration of cooperation in evolutionary dynamics has been a scholarly
pursuit of profound significance, and social dilemmas serve as typical
illustrations of the fundamental conflict between individual and collective
interests in the population. Among the foundational and frequently utilized
models in evolutionary game theory are the prisoner's
dilemma~\cite{axelrod1981evolution} and the snowdrift
game~\cite{smith1997major}, where agents face the binary choices of cooperation
or defection. 

Pioneering work by Nowak and May revealed that the spatial organization of
agents can markedly enhance the emergence of cooperative
behaviors~\cite{nowak1992evolutionary}. However, the study of the snowdrift
game (SDG) showed that spatial structure does not always promote cooperative
behaviors, but often inhibits the evolution of
cooperation~\cite{Hauert2004Spatial}. Subsequent studies discussed the
examination of evolutionary games in various spatial
topologies~\cite{Abramson2000SocialGI,Santos2005ScalefreeNP}. Recently,
numerous mechanisms have been proposed to be effective measures to improve
cooperation in specific situations. Among them, punishment was considered a
viable mechanism for promoting and maintaining
cooperation~\cite{gao2020evolution}. Furthermore, Perc {\it et al.} investigated
various distributions of external factors that influence agent diversity and
concluded that power law distributions are the most conducive to promoting
cooperation~\cite{Perc2007SocialDA}.

A recent study examined the dynamic of information in evolving networks using
the birth-and-death process~\cite{feng2024information}.  Taking into account
the mechanisms of asymmetric interaction, the study conducted by Feng {\it et al.}
elucidated that the presence of a strong detrimental agent can help improve the
incidence of cooperation within the framework of the
SDG~\cite{feng2023harmful}. Li {\it et al.} constructed a resource-based conditional
interaction model where limited resources are redistributed among the
population, and players' resources influence their ability to interact, with a
player paying resources as a learning cost when imitating a neighbor's
strategy~\cite{li2022impact}. A novel game model with heterogeneously
stochastic interactions was proposed, which shows that heterogeneously
stochastic interactions promote cooperation~\cite{li2021evolution}. Moreover,
evolutionary game theory applied to a networked population, along with its
various extensions such as the mixing game and multigame, was empirically
explored to serve as an effective method to address social
dilemmas~\cite{liu2019coevolution,10269140}.

The characteristics of classic networks based on a graph-theoretical framework
were extensively examined in previous studies, leading to the emergence of
numerous complex network models, including multilayer
networks~\cite{wang2015evolutionary,xiong2024coevolution}, temporal
networks~\cite{holme2012temporal,li2017fundamental,zeng2023temporal}, and
higher order networks~\cite{benson2016higher,sheng2024strategy}. These
innovative network frameworks provide a structured foundation for investigating
spatial evolutionary games within populations. Notably, Garde\~nes {\it et al.}
explored evolutionary game dynamics characterized by interdependent networks
reflecting various social ties among individuals~\cite{gomez2012evolution}. In
addition, current research on temporal networks focused primarily on the
dynamics of network edges and cooperation density~\cite{li2020evolution}, with
particular attention to the variability of interaction relationships over
time~\cite{perra2012activity}. Furthermore, the work proposed by
Alvarez-Rodriguez {\it et al.} offered a new method for implementing informed actions
aimed at improving cooperation within social
groups~\cite{alvarez2021evolutionary}. Capraro {\it et al.} explored the shift from
outcome-based to language-based preferences and their impact on moral concerns
and experimental economics~\cite{capraro2024outcome,capraro2024language}.

In real-world interactions, it is observed that most of such engagements are
transient and susceptible to external reorganization, resulting in a dynamic
evolution of social networks over time. Consequently, it is imperative to
recognize that networks represent dynamic entities~\cite{BOCCALETTI2006175}.
The establishment of a dynamic and efficient network highlighted the beneficial
impact of autonomy in fostering cooperative behavior~\cite{su2018promotion}.
Numerous works clarified that cooperation is enhanced when cooperative agents
can achieve a favorable topological positioning. This enhancement is attributed
directly to the avoidance of
defectors~\cite{tanimoto2007dilemma,hanaki2007cooperation}, or indirectly to
the new participants~\cite{Poncela2009EvolutionaryGD}, or ongoing alterations
in the network structure~\cite{szolnoki2008making}. Numerous research studies
have been conducted on the adjustment of social ties for individuals to select
their peers based on local reputation
information~\cite{Fu2008ReputationbasedPC,2019Reputation}. Szolnoki {\it et al.}
demonstrated that coevolving random networks can develop appropriate mechanisms
for each social dilemma, enabling cooperation to thrive even in adverse
conditions~\cite{szolnoki2009resolving}. Taking into account the arbitrary
spatial and temporal heterogeneity, a recent study showed that transitions
among a large class of network structures favor the spread of cooperation, even
if each social network would inhibit cooperation when
static~\cite{su2023strategy}. Furthermore, asymmetrical interactions have
emerged as a critical focal point in contemporary research, warranting
significant attention and investigation~\cite{li2024asymmetrical}.

More recently, there has been an increasing focus on the dynamics of
adaptive~\cite{gross2008adaptive} or coevolutionary
networks~\cite{perc2010coevolutionary}, where connectivity is modified
dynamically corresponding to their dynamical states. A series of scholarly
investigations investigated how agents are capable of modulating their
interactions with neighboring entities, employing withdrawal or maintenance
strategies, depending on their level of satisfaction and selected
approach~\cite{Pacheco2006CoevolutionOS,Zschaler2009AHR}. Research has shown
that the coevolutionary rule which affects the random topology of the
interaction network triggers the spontaneous emergence of a robust multilevel
selection mechanism~\cite{Szolnoki2009EmergenceOM}. In addition, Zimmermann
{\it et al.} explored the dynamic relationship between the internal states of the
interacting elements and their interaction
network~\cite{Zimmermann2004CoevolutionOD} and examined scenarios where the
network locally adjusts based on the overall payoff of the
agents~\cite{Zimmermann2000CooperationIA}. Yao {\it et al.} investigated
co-evolutionary dynamics of cooperation by examining how willingness to
participate in social interactions and payoff accumulation influence the
activation or inhibition of network edges, ultimately affecting local network
dynamics and the evolution of cooperation under varying sensitivities and
temptations to defect~\cite{yao2023inhibition}. As a further step, a new
network adaptation by separating disassortative neighbors was proven to be an
effective way to resolve the social dilemma~\cite{MIYAJI2021110603}.

By following the previous path, we primarily focus on investigating the adaptive evolving network in which the network structure jointly evolves with the states of the interacting agents. Within such a framework, the population structure can be captured by a dynamic network, wherein each node signifies agents, while the edges symbolize their alterable interactions. Agents probably imitate current neighbors who have higher payoffs over time, and it is common experience that agents tend to cut interaction links when faced with adverse neighborhoods. As a reasonable assumption, it can be posited that, after altering the strategies, the disassociation tendency between any pair of neighbors is asymmetric and depends on the comparative payoff that the focal agent derives from all neighbors. More precisely, the lower the payoff that a specific neighbor brings to the focal agent, the higher the tendency of the focal agent to dissociate from that neighbor, which directly impacts whether they will persist as neighbors in the next step, i.e.,~determining the persistence of the edge between them. In addition, agents who terminate interactions with unfavorable neighbors probably choose new neighbors to replace them. 

The proposed adaptive mechanism allows for the exploration of diverse strategies and facilitates the emergence of novel interaction patterns within the social system. We experimentally investigate the time steps required for a network to reach a stationary state under various initial configurations, as well as the proportion of cooperators within the network. These results are compared with those obtained from static networks with identical configurations. Subsequently, we analyze the changes in different types of edges during the evolutionary process. Furthermore, we elucidate the degree distribution of all agents upon reaching the stationary state, distinguishing between the degree distributions of cooperators and defectors. Finally, the average clustering coefficient and average path length of the network are also examined at both the initial stage and after evolving to the stationary state.

The paper is organized as follows. In \ref{sec:model}, we delineate the detailed rules that govern both strategy and network evolution. The key results and significant experimental findings are presented in \ref{sec:results}. Finally, \ref{sec:conclusion} offers a summary of our findings and a discussion on potential directions for future research improvements.

\section{Model}
\label{sec:model}

Interaction dynamics is often characterized by the possibility of severing existing edges and forming new ones. Within the framework of game theory, these dynamic changes in interactions are primarily influenced by the payoffs received from current neighbors. Motivated by this reality, we propose an adaptive and coevolutionary model for the SDG. We consider each agent to have the capability to sever certain edges at the end of each round of the game. At each time step, every agent exhibits a specific disassociation tendency to all of its neighbors based on the received payoffs. In addition, agents who opt to terminate interactions with unfavorable neighbors have the freedom to choose new neighbors to replace them. Based on these assumptions, we introduce the SDG model along with delineating agents' payoffs and the strategy update rules. Subsequently, we define the concept of an asymmetric disassociation tendency, which serves as the crucial determinant for the probability of adjusting unfavorable connections. Therefore, our proposed adaptive network structure not only allows for the termination of existing connections but also accommodates the establishment of new ones, which ensures that the aspect provides a comprehensive exploration of microscopic dynamics, incorporating both the dissolution and formation of connections.

 \subsection{Strategy evolution}

The classic SDG framework encapsulates a scenario in which two drivers find their way home obstructed by a snowdrift. In this social dilemma, each agent faces the decision of cooperating ($C$) or defecting ($D$). When both parties opt for cooperation, it enables both to reach their common goal, resulting in an equivalent return, denoted $R=1$. In contrast, if both agents decide to defect, their progress is impeded, leading to no return $P=0$. Alternatively, if an agent chooses the defection strategy and encounters a neighbor who has chosen cooperation, they can ultimately reach their destination. It is essential to note that in this situation, the agent opting to defect attains the maximum return without any cost represented as $T=1+r$ (where $0<r<1$). Although the cooperative agent receives a lesser return, denoted as $S=1-r$, where $r$ represents the cost-to-benefit ratio. Consequently, the resultant payoff matrix $\mathbf{A}$ can be expressed as in~\eqref{eq:1}, satisfying $T > R > S > P$ and $2R = S + T$.
\begin{equation}
\label{eq:1}
\mathbf{A}=
\begin{pmatrix}
R & S \\
T & P 
\end{pmatrix}
=
\begin{pmatrix}
1 & 1 - r \\
1 + r & 0 
\end{pmatrix}
.
\end{equation}

Consider a population consisting of $N$ agents, each participating in pairwise SDG interactions with their neighbors. In this context, the strategic choices available to each agent $s_i$ are defined as cooperation denoted as $[1,0]$, or defection termed as $[0,1]$. In this way, the neighborhood of any given agent $i$, denoted as $\mathcal{N}_i$, comprises agents directly interconnected to agent $i$ through a single edge. The size of $\mathcal{N}_i$ defines the degree of $i$, aptly termed $k_i$. At each time step, every agent engages in interactions with all neighbors and subsequently accumulates its total payoff by~\eqref{eq:2}.

\begin{equation}
\label{eq:2}
\Pi_i=\sum_{j \in \mathcal{N}_i} \Pi_{ij}=\sum_{j \in \mathcal{N}_i} s_i \mathbf{A} s_j^T
,
\end{equation}
where $i$ denotes any neighboring agent of agent $j$, and $\Pi_{ij}$ is the corresponding payoff between them.

After interacting with all neighbors in each round, all agents have the opportunity to update their strategies. Here, an agent $i$ randomly chooses one of its neighbors and imitates the strategy of $j$ with a probability defined by Equation~\ref{eq:3}, meaning that the agent is restricted to imitate the strategy of its neighbor who attains higher payoffs.

\begin{equation}
\label{eq:3}
P(\left.S_i \rightarrow S_j\right)= \begin{cases}\frac{\Pi_i-\Pi_j}{(1+r) \cdot \max \left(k_i, k_j\right)} & , \Pi_i \geqslant \Pi_j \\
0 & , \Pi_i<\Pi_j\end{cases}.
\end{equation}

\subsection{Network evolution}
 
Agents are disinclined to interact with neighbors who produce lower rewards for themselves. In other words, agents develop a disassociation tendency based on their own benefits in relation to all neighbors at each time step. Therefore, we assess the disassociation tendency of the agent $i$ toward the neighbor $j$ at time $t$ according to the dynamics defined by Equation~\ref{eq:4}:

\begin{equation}
\label{eq:4}
D_{i j}(t) = \frac{1}{k_i} \sum_{\substack{m \in \mathcal{N}_{i}(t)}} [\Pi_{i m}(t) - \Pi_{i j}(t)]
,
\end{equation}
where $m$ represents every neighbor of agent $i$. $\Pi_{ij}(t)$ denotes the payoff that neighbor $j$ provides to agent $i$ at time $t$, and similarly $\Pi_{im}(t)$ is the payoff contributed to $i$ by all its neighbors.

\begin{figure}
	\centering 
	\includegraphics[width=\columnwidth]{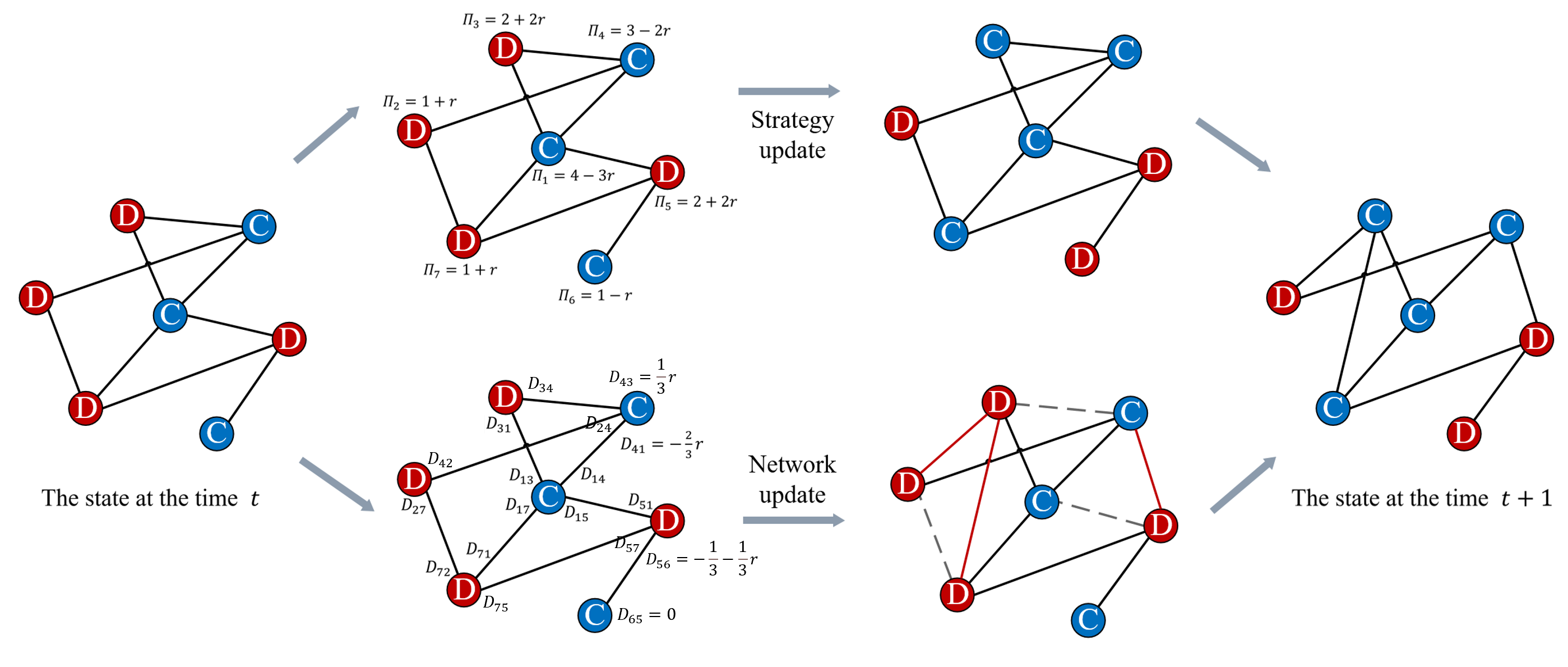}	
	\caption{\textbf{A schematic representation of the adaptive mechanism: evolution of network state from $t$ to $t+1$.} Here, cooperative nodes are visually denoted by the color blue, whereas defectors are represented in red. We partition the evolutionary process into two distinct phases: the strategy updating and the network updating, elucidating its underlying principles and mechanisms. During the strategy updating, agents adjust their strategies based on the accrued payoffs $\pi_{ij}$, while in the network updating, agents rewire their connections driven by the disassociation tendency $D_{ij}$. In the network updating phase, gray dashed lines indicate the connections to be severed, while red lines represent newly formed connections following the adjustment of adverse neighbor relationships.} 
 \label{fig:model}
\end{figure}

As a crucial element of our model, we assume that the disassociation tendency between a pair of neighbors $i$ and $j$ is asymmetric, i.e., $D_{ij}(t)$ is not necessarily equal $D_{ji}(t)$. It is worth highlighting that both $D_{ij}(t)$ and $D_{ji}(t)$ collectively influence whether they maintain their neighborly relationship at time $t+1$. Consequently, we define the probability $\Gamma_{ij}(t+1)$ that the edge between neighbors $i$ and $j$ will not be connected in the next time step as in Equation~\ref{eq:5}, ensuring that the probability always remains in the $[0,1]$ interval.

\begin{equation}
\label{eq:5}
\Gamma_{ij}(t+1) = \frac{1}{1 + e^{-\frac{D_{ij}(t) + D_{ji}(t)}{2}}}.
\end{equation}

Generally, agents tend to look for new neighbors after adjusting the adverse social links. When an agent is untied to a neighbor, we provide the agent with the autonomy to select a new neighbor or maintain its current set of connections without forming a new link. Formally, we consider two nodes $i$ and $j$ that are neighbors at time $t$. If nodes $i$ and $j$ decide to no longer interact at time $t+1$, they both possess the autonomy to choose new connections. The process ensures the stability of the total number of edges in the network, as the autonomy granted to the agents in selecting new neighbors or retaining the current state prevents abrupt changes in the network's overall connectivity.

To encapsulate our model, Fig.~\ref{fig:model} depicts a prototypical adaptive mechanism where the state of the network evolves over time $t$ to $t+1$. Regarding the strategy update, agents probably imitate their neighbors with higher pay-offs through pairwise comparison. In addition to them, updates to the network structure allow agents to cut off ties with unfavorable neighbors and autonomously seek new connections. Specifically, each pair of neighbors possesses a set of asymmetric dissociation tendencies, influenced by their relative payoffs, which reflects the probability of them not remaining neighbors in the subsequent time step. For instance, Agent 4 exhibits different dissociation tendencies toward its neighbors, Agents 1 and 3. It is illustrated in the diagram where $D_{43}= \frac{1}{3}r$ and $D_{41}=-\frac{2}{3}r$, which are derived from Equation~\ref{eq:4} and depend on the specific payoff contributions of each neighbor. Additionally, Agents 5 and 6 exhibit asymmetric dissociation tendencies toward each other, stemming from the contrasting nature of their relationships within the neighborhood network. It is noteworthy that when pairs of agents contemplate the termination of their connections in the subsequent time step, three distinct scenarios might unfold: first, both parties opt to seek new connections independently; secondly, one party seeks new connections while the other does not; or thirdly, neither party establishes new connections.

The intertwined coevolution of strategy and network connection underscores the adaptive nature of the system, wherein agents continually adjust their strategies based on the success of their neighbors while simultaneously reshaping their social network connections. This symbiotic relationship between strategy and network dynamics not only influences individual agent behavior but also shapes the emergent properties and collective behavior of the entire network over time.

\section{Analysis of the simulation results}
\label{sec:results}
In this section, we describe our simulation methodology and present the findings that illustrate the coevolution of cooperation levels and network structures in various parameter configurations. To elucidate the potential influence of various initial network topologies on system dynamics, we employ two distinct network models: random regular graph (RG)~\cite{bollobas1982diameter} and Watts-Strogatz small world network (WS)~\cite{watts1998collective}. WS is generated with $p_r = 0.1$ as its rewiring probability. Furthermore, both RG and WS networks with average degrees of $k = 4$ and $k = 10$. In all cases, the complete network comprises $N = 2,000$ nodes. Initially, each agent is randomly designated as either a cooperator or a defector with an equal probability. Simulations are programmed in Python 3.9 using the Anaconda environment.

\subsection{Influence of the adaptive network on the cooperation}

\begin{figure*}  
    \centering
    \hspace{-5.6mm}
    \subfigure[RG]
    {
    \includegraphics[scale=0.47]{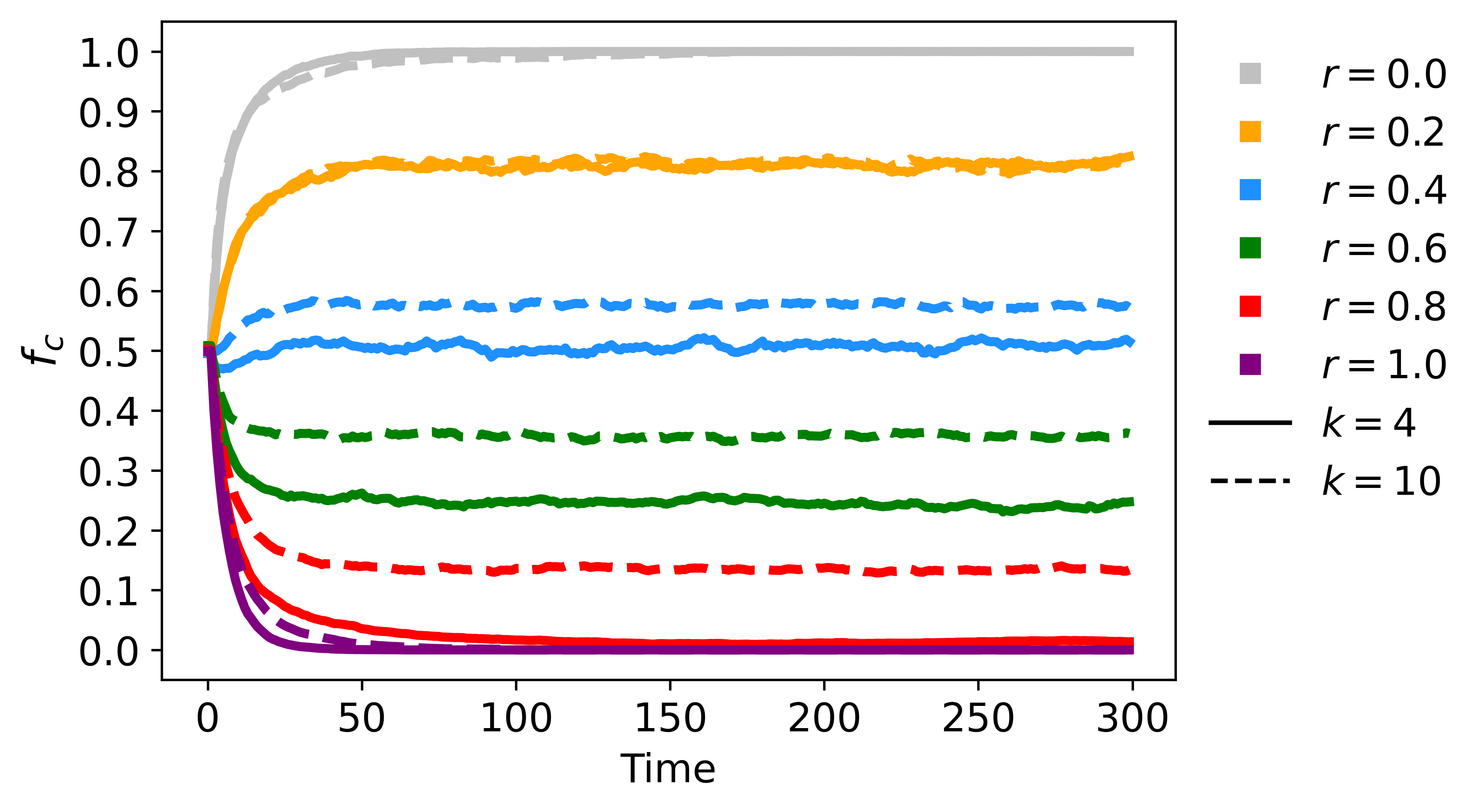}
    }
    \hspace{-1mm}
    \subfigure[WS]
    {
    \includegraphics[scale=0.47]{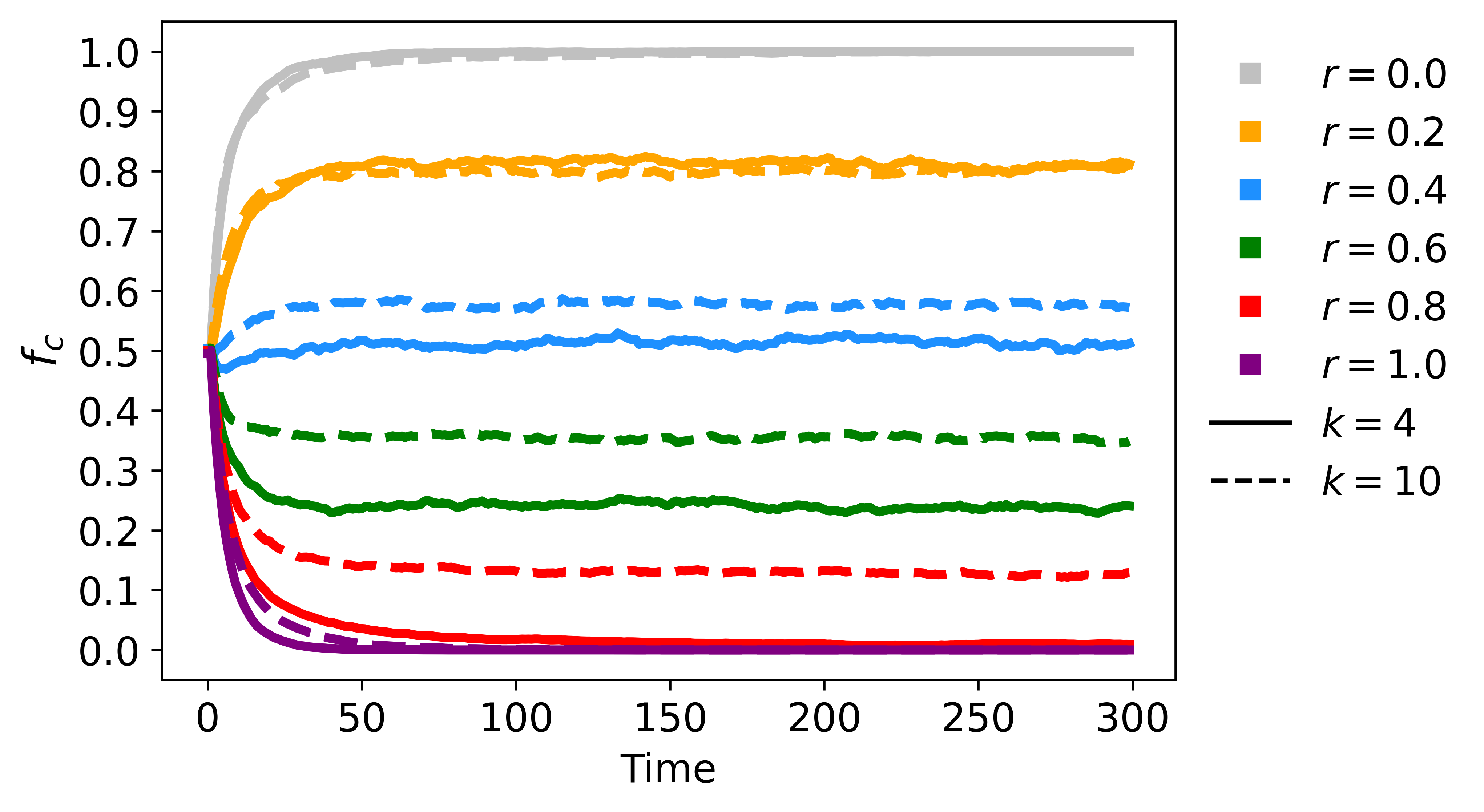}
    }
    \hspace{-0.2mm}
    \caption{\textbf{Time evolution of $f_c$ for various $r$ values under different initial network configurations.} The evolution of $f_c$ is examined for various values of $r$ with respect to the number of evolution iterations depicted for initial network structures of RG (a) and WS (b), where solid and dashed lines correspond to $k=4$ and $k=10$ respectively. Different colors in the curves indicate distinct values of $r$ as illustrated in the legend. Across all initial topologies, cooperation flourishes predominantly at lower values of $r$, with evolution stabilizing at approximately 200 steps.}
    \label{fig:2}
\end{figure*}

First, we examine the evolutionary dynamics of the fraction of cooperators ($f_c$) across various types and average degree networks under varying $r$, which involves an investigation of the evolutionary process of $f_c$ as it progresses toward a stationary state under different parameter conditions over time. Fig.~\ref{fig:2} illustrates the average alterations of $f_c$ through 10 independent runs with different parameter values over iterations. Any initial network structure is more likely to evolve to a stationary state at a slower pace compared to traditional static models because of the dynamic adaptation of the network structure itself. 

However, as illustrated in Fig.~\ref{fig:2}, for any given network structure and parameter conditions set here, $f_c$ consistently converges to a stationary state within 200 iterations. For both initial RG (a) and WS (b) structures, their impact on the temporal evolution of $f_c$ is not significant. It demonstrates the resilience of the adaptive network model, which can stabilize cooperative behavior even with different initial network structures like RG or WS. Generally, lower values of $r$ are associated with a higher prevalence of cooperation, while larger values correspond to an increased presence of defection. Furthermore, as $r$ approaches the extreme (0 or 1), the evolutionary dynamics of $f_c$ demonstrate a heightened pace, indicating a more rapid transition in the prevalence of cooperation. Moreover, under extreme values of $r$, the influence of initial network structure and different average degree values on the evolution of cooperation can be completely ignored, as prominently evidenced even in $r=0.2$. 

Fig.~\ref{fig:2} also reveals how different network structures and average degrees influence cooperation dynamics. It is notable that under other conditions ($r=0.4$, $0.6$, $0.8$), the networks of $k=10$ exhibit a significantly higher level of cooperation compared to $k=4$, implying that having more neighbors to evaluate the tendency to disassociation is more conducive to the prevalence of cooperation. Especially noteworthy is the case where $r=0.8$, where the cooperation frequency eventually decreases to 0 in the networks of $k=4$, while a higher average degree, $k=10$, remarkably enhances $f_c$ and accelerates its attainment of a stationary state.

\begin{figure*}  
    \centering  
    \subfigure[RG]
    {
    \includegraphics[scale=0.067]{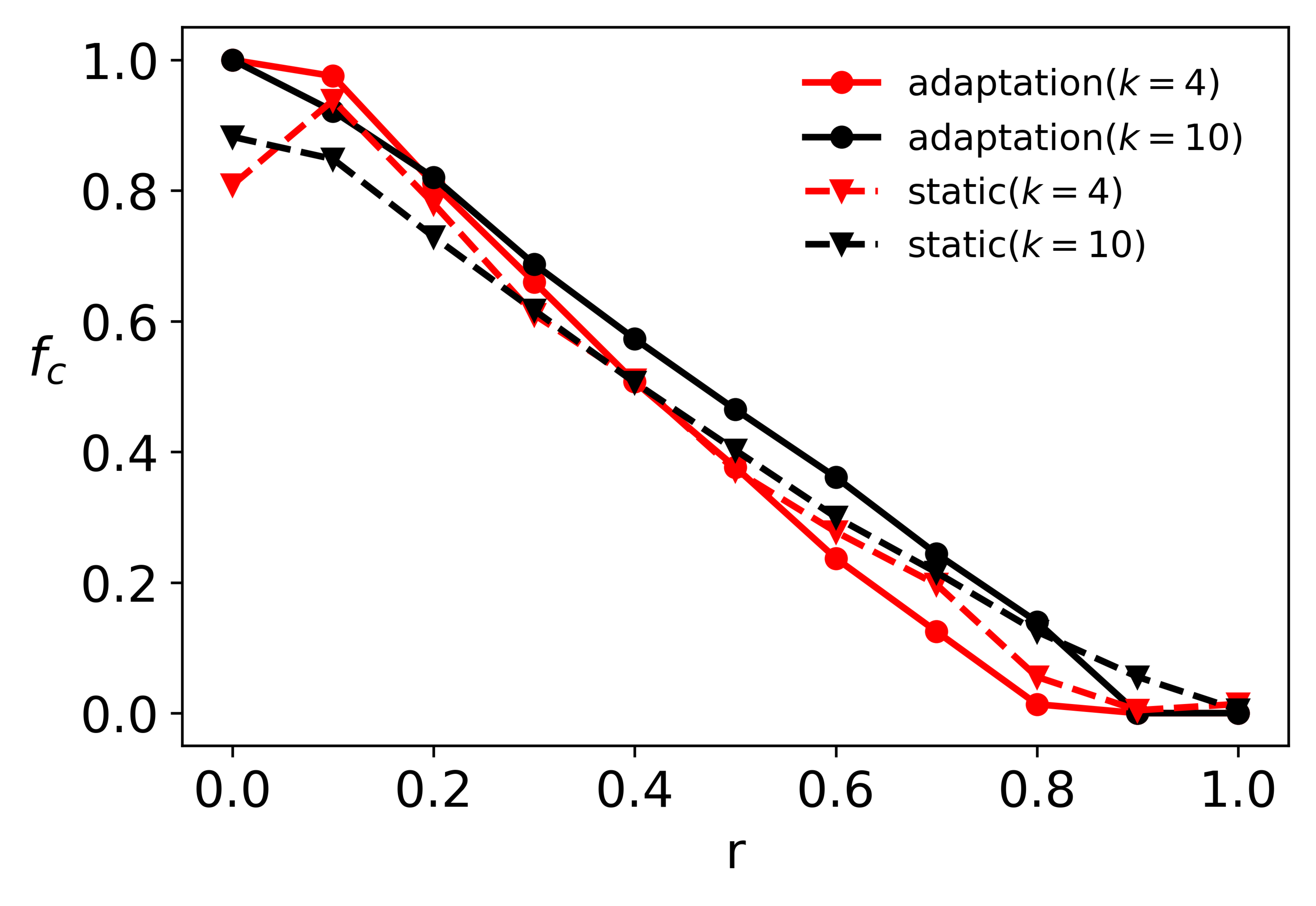}
    \label{fig3:RG}
    }
    \hspace{-0.2mm}
    \subfigure[WS]
    {
    \includegraphics[scale=0.067]{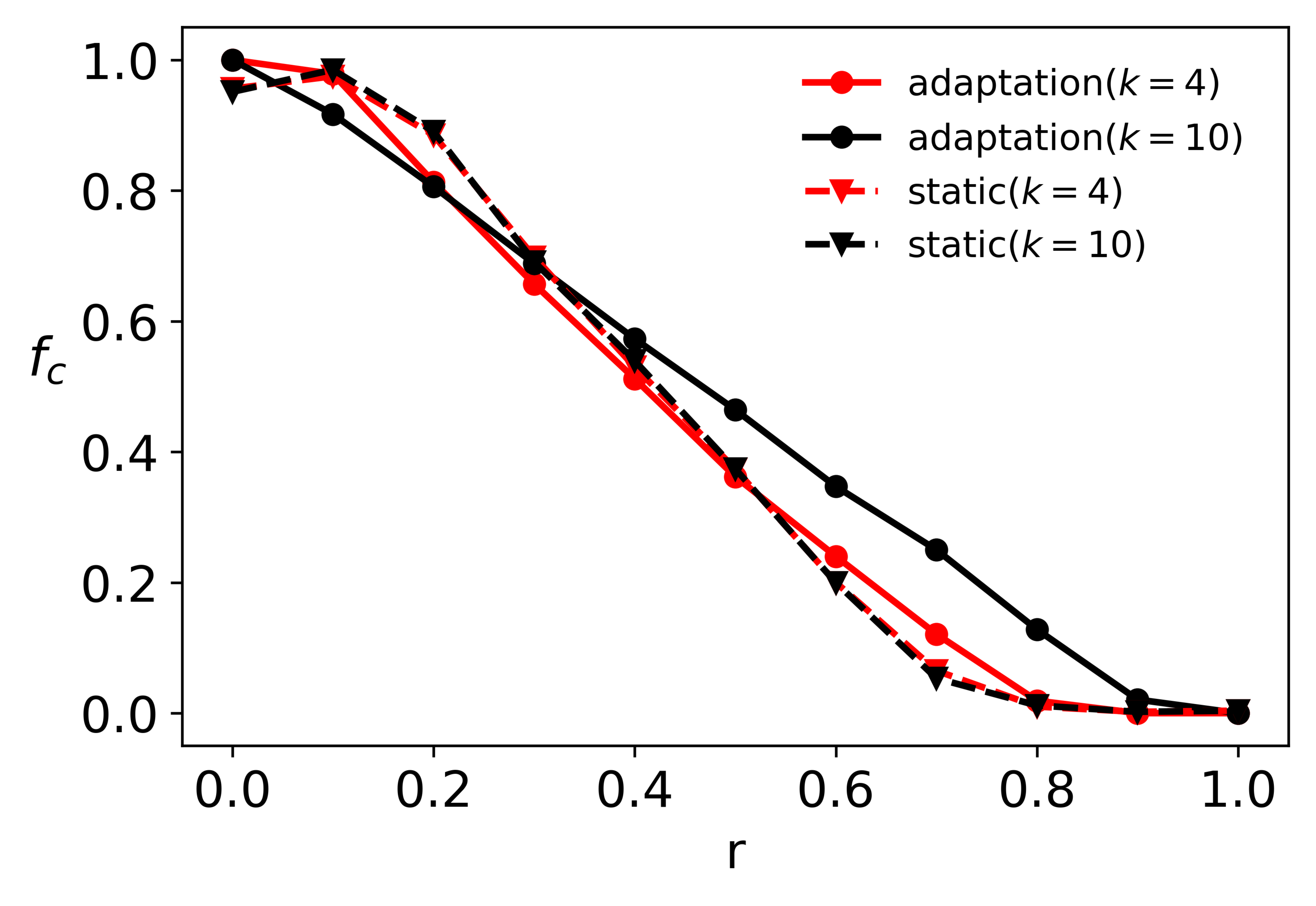}
    \label{fig3:WS}
    }
    \hspace{-0.2mm}
    \caption{\textbf{The $f_c$ is investigated as a function of $r$ ranging from 0 to 1.0 for both adaptive and static networks.} It is explored how $f_c$ varies with $r$ under a consistent initial network structure. Essentially, the initial topological structure of the adaptive network mirrors that of the static network, yet the adaptive network undergoes experiences co-evolution according to the pre-defined adaptive rules. Significant variability in $f_c$ between different $r$ values and network topologies is observed, with distinct trends emerging at lower and higher $r$  values, which highlights the enhanced adaptability of dynamic networks to foster cooperation under varying environmental constraints compared to rigid static networks.}
    \label{fig:3}
\end{figure*}

The adaptability allows the system to self-organize and optimize connections, creating advantageous cooperative clusters that traditional static networks struggle to replicate. To further investigate the discrepancies between our proposed adaptive networks and traditional static networks based on the same game model, we analyze the stability of $f_c$ over evolutionary iterations, as shown in Fig.~\ref{fig:3}. It is essential to highlight that the adaptive co-evolving network shares identical initial network types and average degrees with the static network configuration. In contrast to static networks, our adaptive network exhibits a propensity toward evolving into either pure cooperation or pure defection states under sufficiently small or large values of $r$, a trend that is scarcely observed in static networks. 

Specifically, the function $f_c$ of the RG ($k=4$) topology reveals that within the parameter range where $r \leq 0.5$, the adaptive network exhibits a greater proportion of cooperators compared to instances where $r$ exceeds 0.5. Within the broader range of $0 \leq r \leq 0.8$, the density of cooperation of the adaptive RG with $k=10$ consistently outperforms that of the static network with the same topology, while as $r$ increases, the static network demonstrates a relatively higher level of cooperation compared to the adaptive network. Interestingly, the $f_c$ values for static WS networks with $k=4$ and $k=10$ exhibit negligible disparities, with a sudden spike observed at $r=0.1$. In contrast, adaptive networks manifest significantly divergent system behaviors in various average degree configurations.

In the case of WS networks with $k=4$, both dynamic and static networks display advantages at different $r$ values, although differences are not significant. However, for adaptive networks with WS topology and $k=10$, a noticeable advantage emerges under parameter conditions where $r$ is greater than or equal to 0.4. When the value of $r$ is relatively high, indicating that defectors dominate the system, we observe that the $f_c$ in adaptive networks is higher compared to that in static networks. Based on this, we can infer that, as numerous previous research works have highlighted, the essential function of network configurations in cultivating cooperative clusters and consequently defending against incursions by defectors is significant. Dynamic adaptive networks indeed exhibit distinct characteristics in system behavior compared to traditional fixed network structures. To validate the formation of these cooperative clusters, we next further analyze this dynamic network topology.

\subsection{Impact of the dynamic network topology}
\begin{figure*}  
    \centering
    
    \subfigure[]
    {
    \includegraphics[scale=0.634]{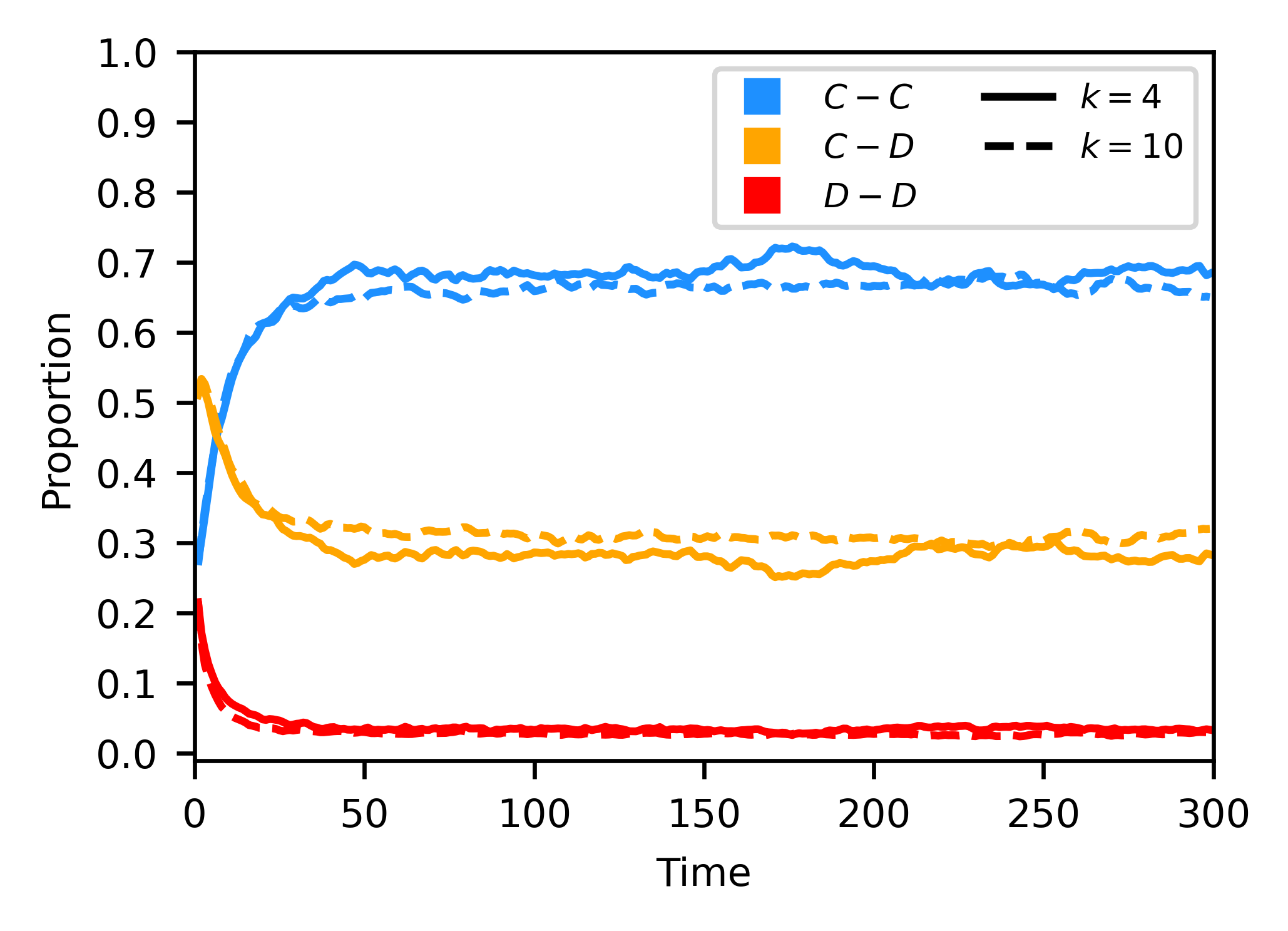}
    \label{fig4:RG-0.2}
    }
    \hspace{-5.99mm}
    \subfigure[]
    {
    \includegraphics[scale=0.634]{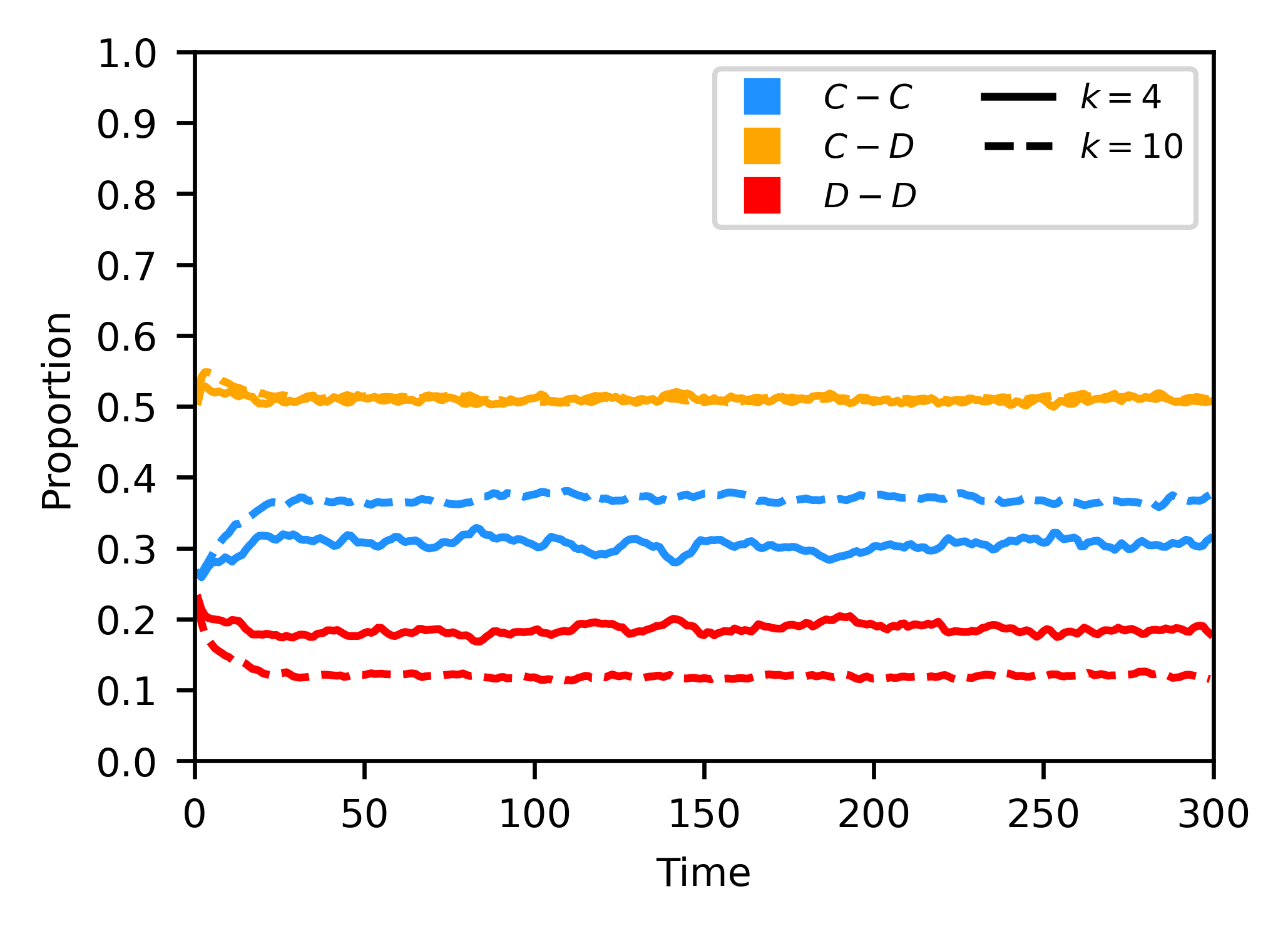}
    \label{fig4:RG-0.4}
    }
    \hspace{-5.99mm}
    \subfigure[]
    {
    \includegraphics[scale=0.634]{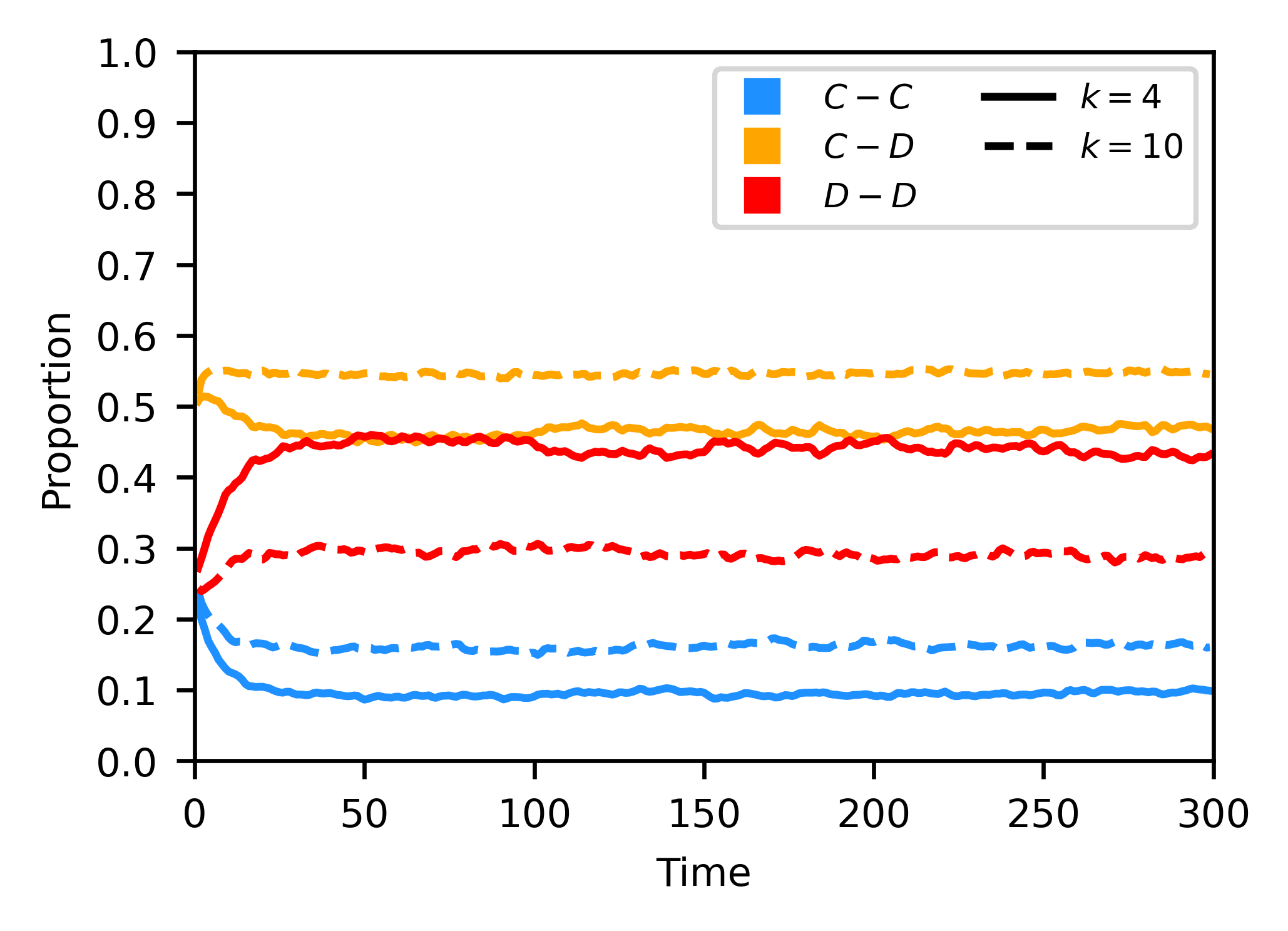}
    \label{fig4:RG-0.6}
    }
     
    \subfigure[]
    {
    \includegraphics[scale=0.63]{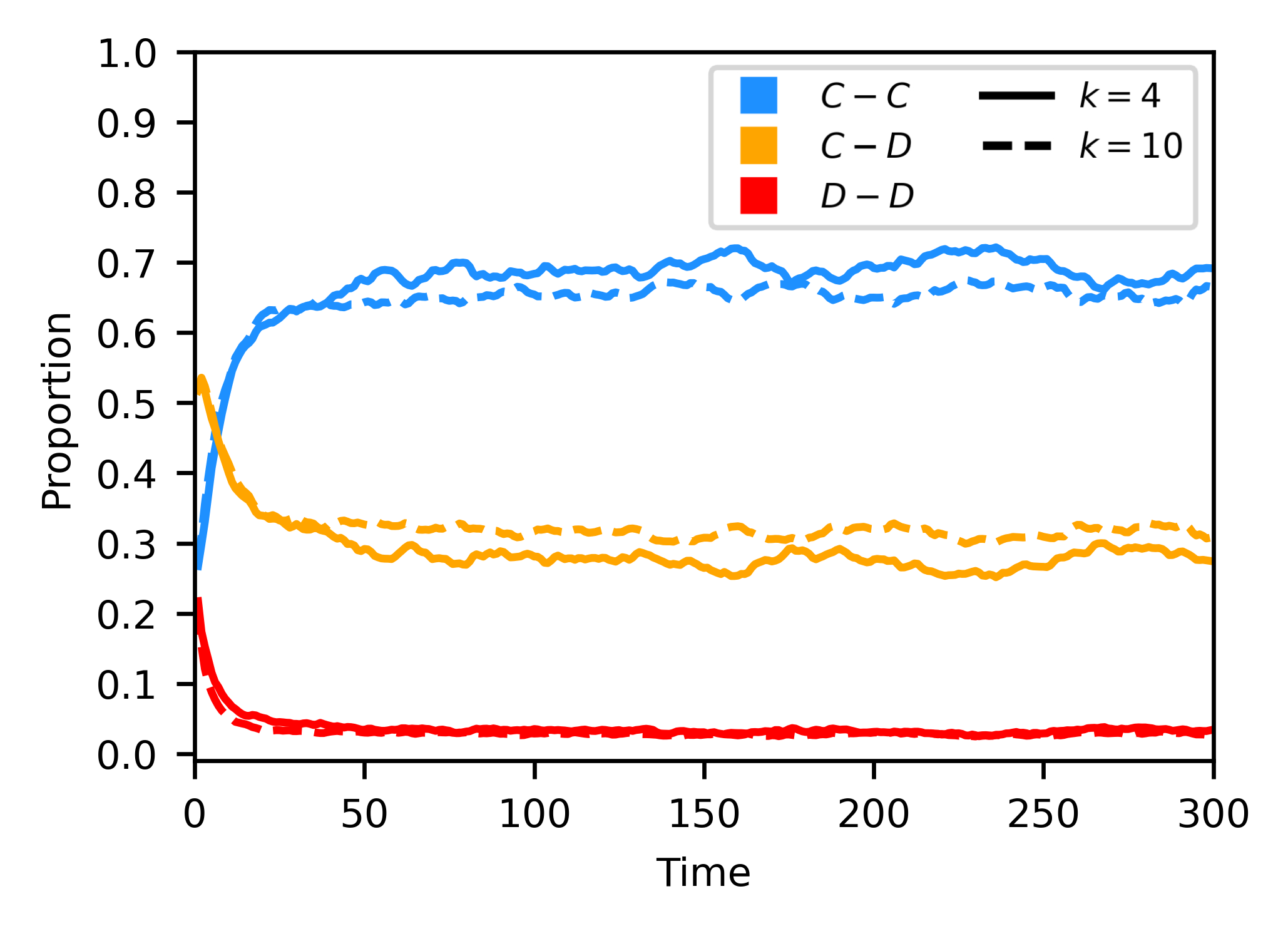}
    }
    \hspace{-5.99mm}
    \subfigure[]
    {
    \includegraphics[scale=0.632]{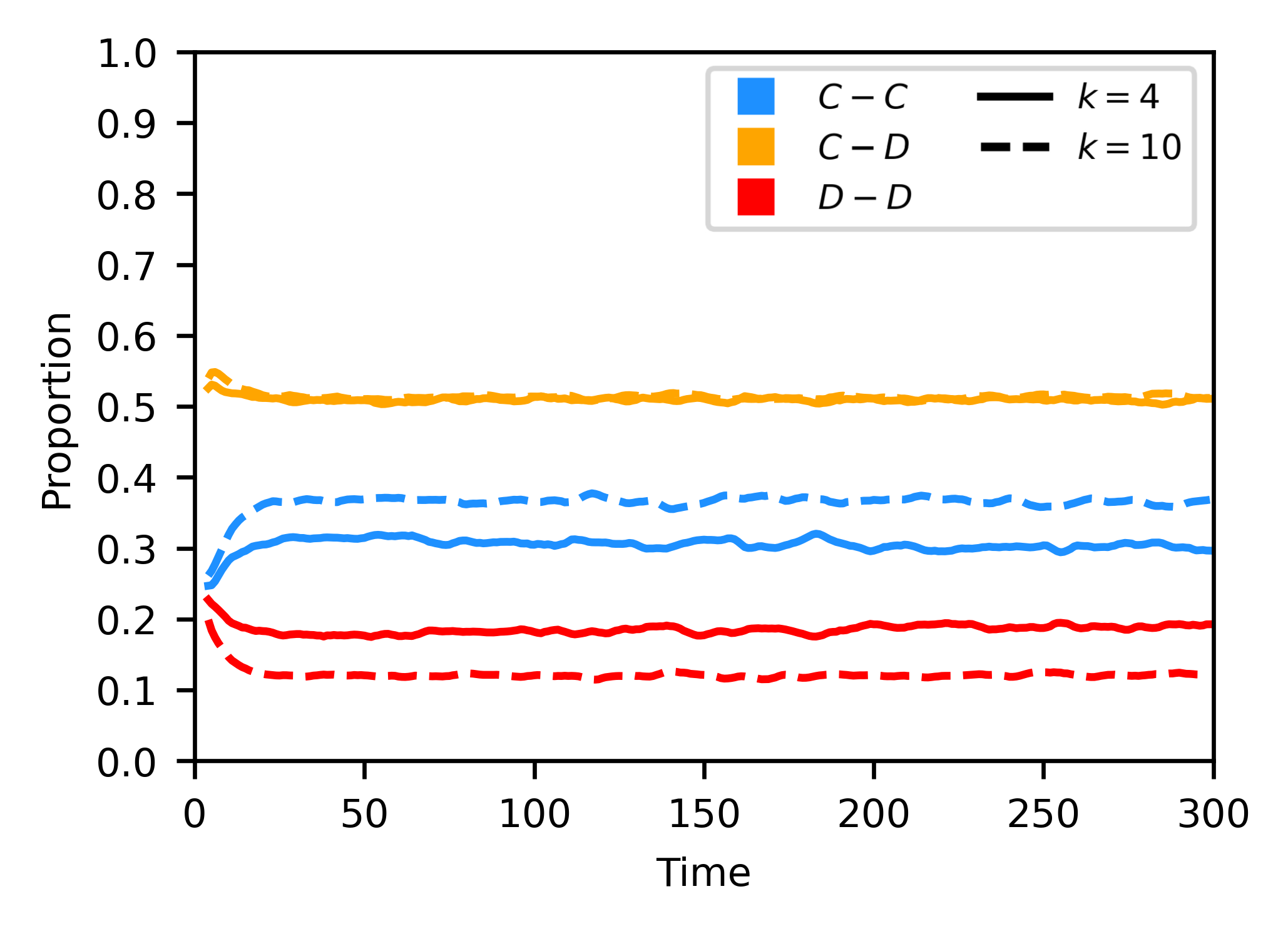}
    }
    \hspace{-5.99mm}
    \subfigure[]
    {
    \includegraphics[scale=0.632]{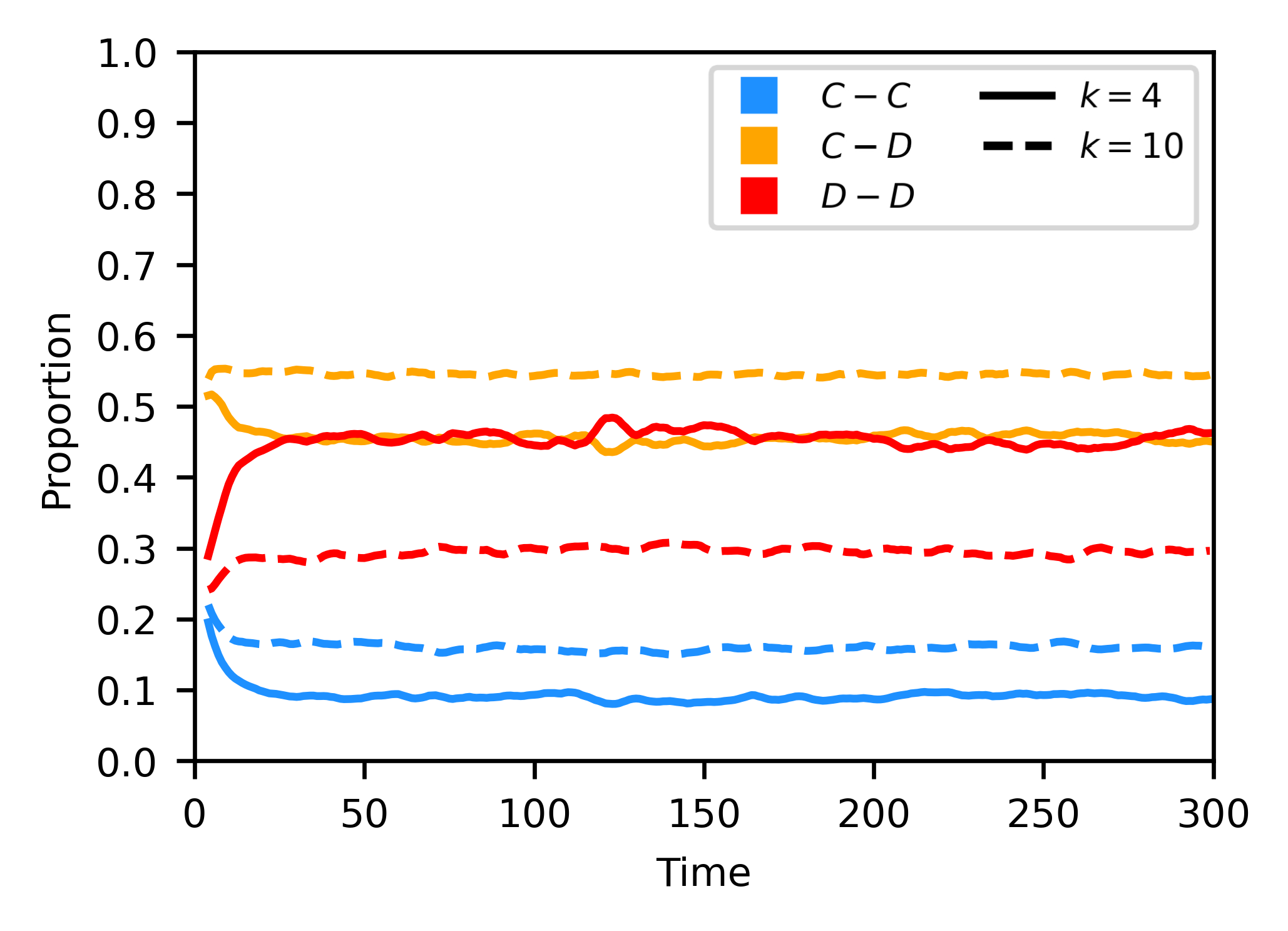}
    }
    \caption{\textbf{The proportions of C-C, C-D, and D-D edge types vary across different networks and $r$ values.} The upper row of the network structures depicts RG for all cases across various $r$ values: $r = 0.2$ (a), 0.4 (b), and 0.6 (c). Consequently, the lower row applies WS with the same values for $r$ as the upper row: $r = 0.2$ (d), 0.4 (e), and 0.6 (f). The solid lines represent an average degree of the network of $k=4$, while the dashed lines represent $k=10$. Blue, orange, and red represent the C-C, C-D, and D-D edge types, respectively. Upon contrasting the two sets of sub-figures, the initial network structures exhibit an insignificant influence on edge types. The first column indicates that, under a scenario dominated by cooperation, the C-C type edges increase rapidly, while the proportion of D-D type edges decreases faster than that of C-D type edges. Results from the second column suggest that when the ratios of cooperation and defection are similar, the proportion of C-D type edges remains almost constant, with a slight increase in C-C edges and a corresponding decrease in D-D edges. The third column implies that, under a predominance of defectors, there is an increase in the proportion of D-D edges, which become nearly equal to C-D type edges, and noticeably, networks with $k=4$ and $k=10$ exhibit significant differences.}
    \label{fig:4}
\end{figure*}

\begin{figure*}  
    \centering
    \subfigure[]
    {
    \includegraphics[scale=0.46]{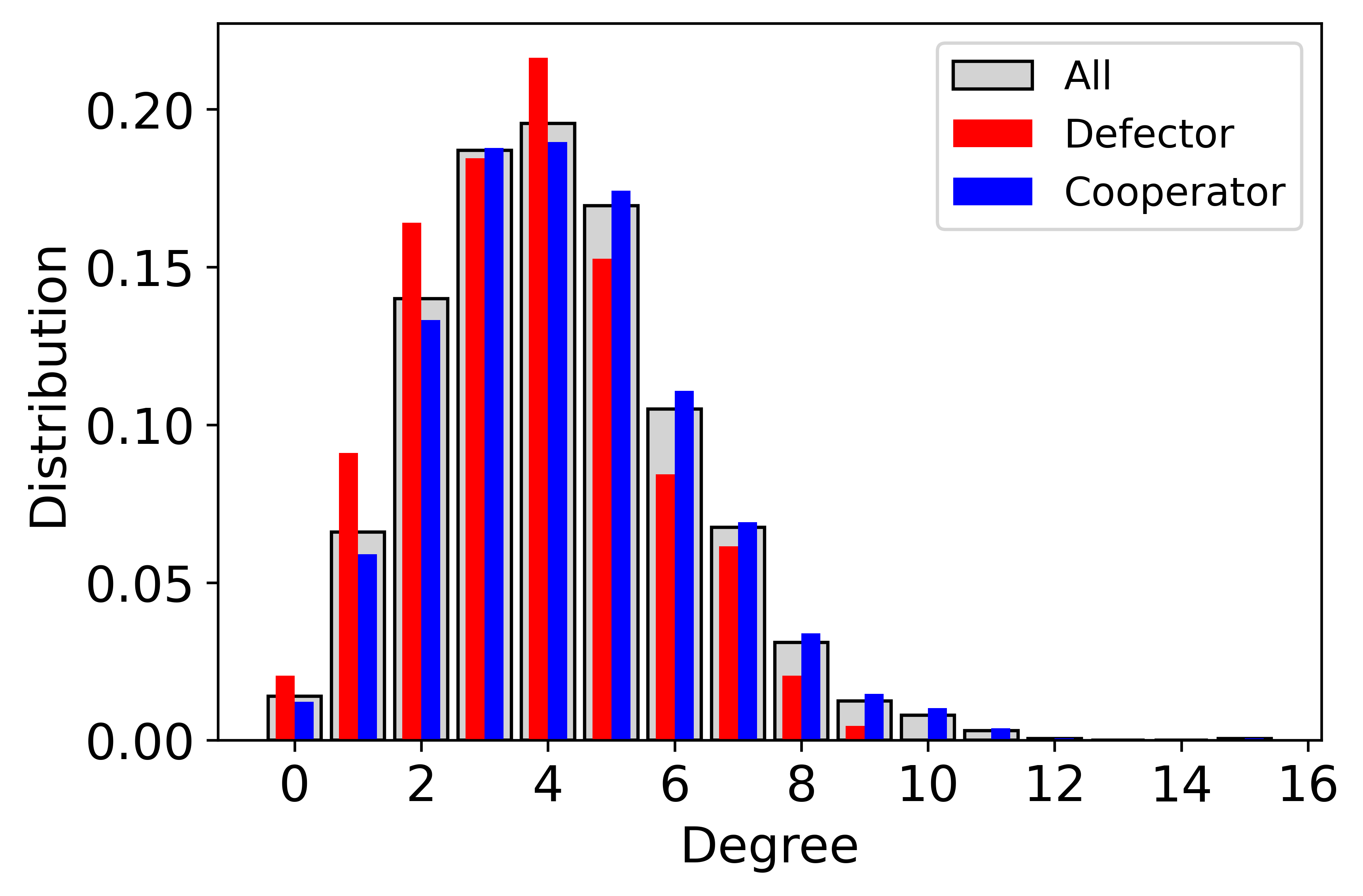}
    } 
    \subfigure[]
    {
    \includegraphics[scale=0.46]{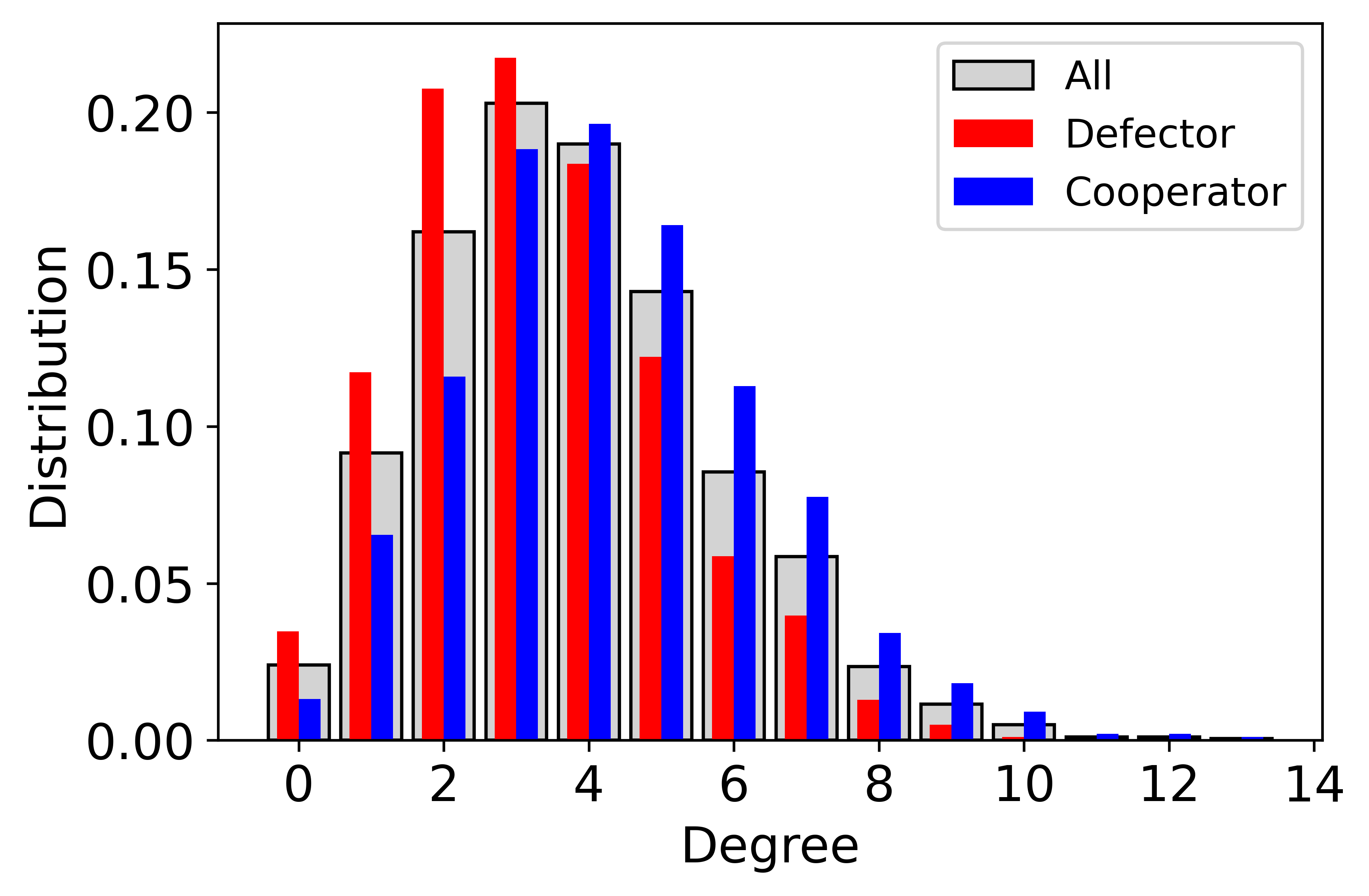}
    }
    \subfigure[]
    {
    \includegraphics[scale=0.46]{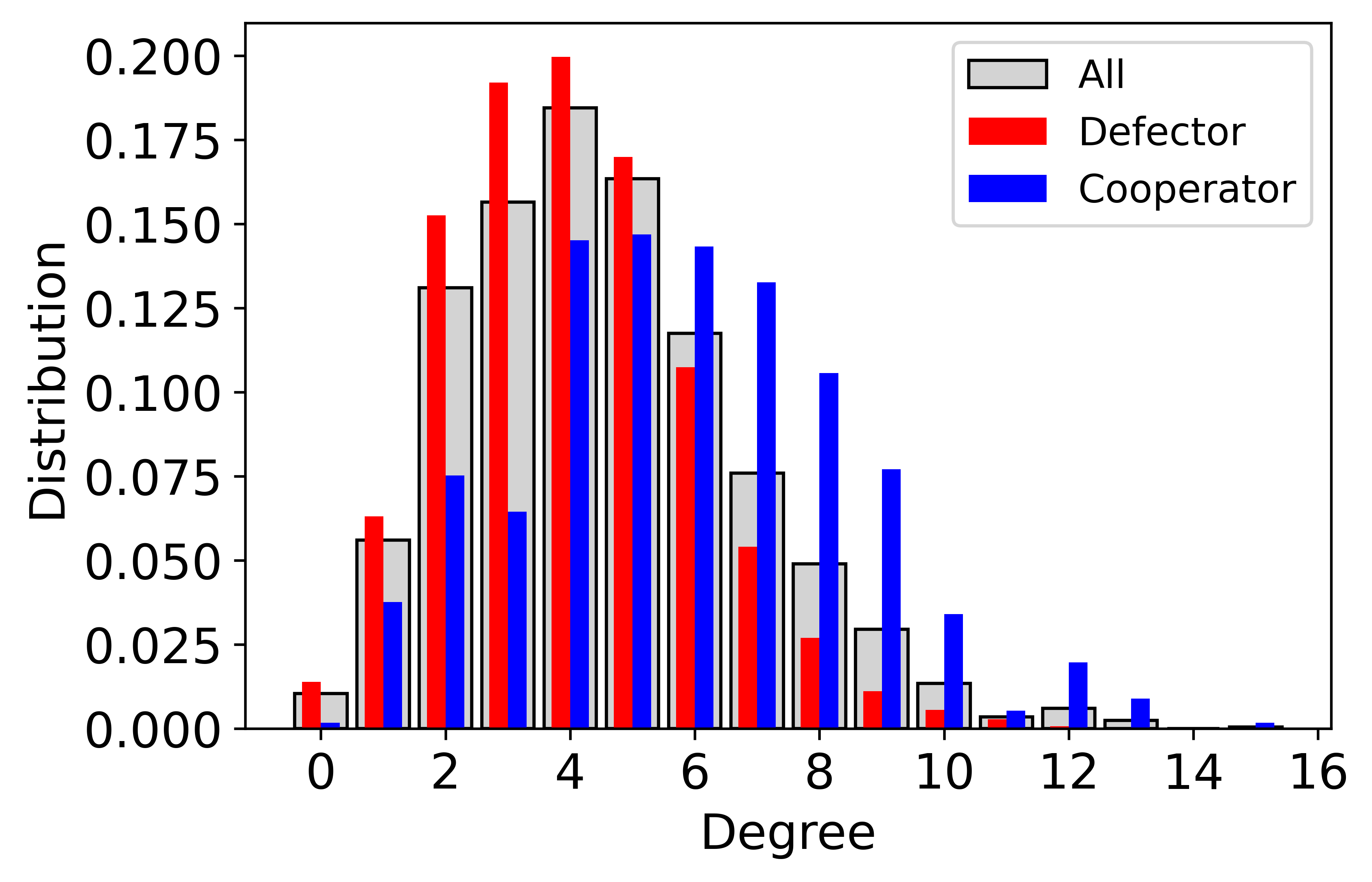}
    }       
    \subfigure[]
    {
    \includegraphics[scale=0.46]{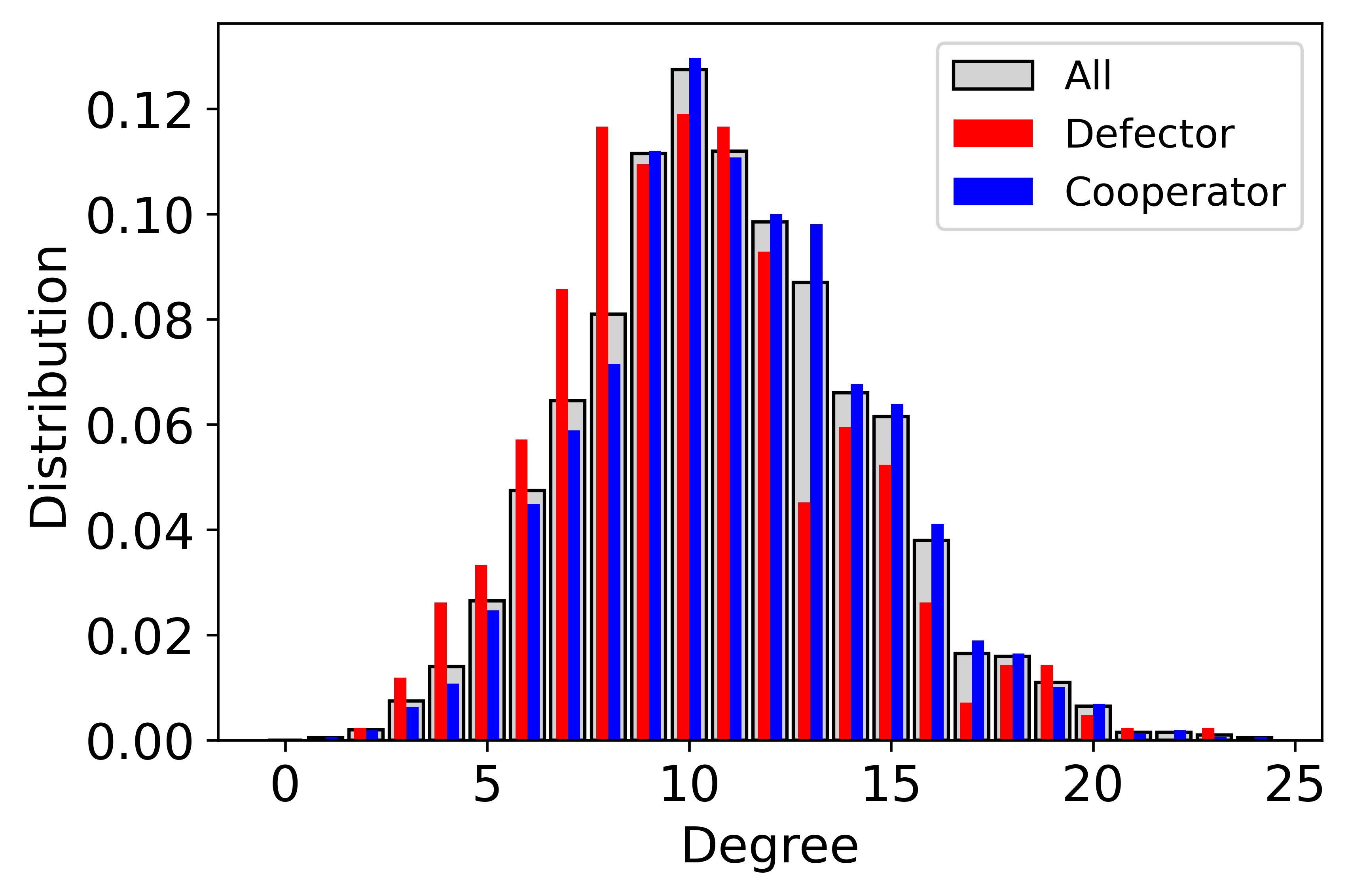}
    }
    \subfigure[]
    {
    \includegraphics[scale=0.46]{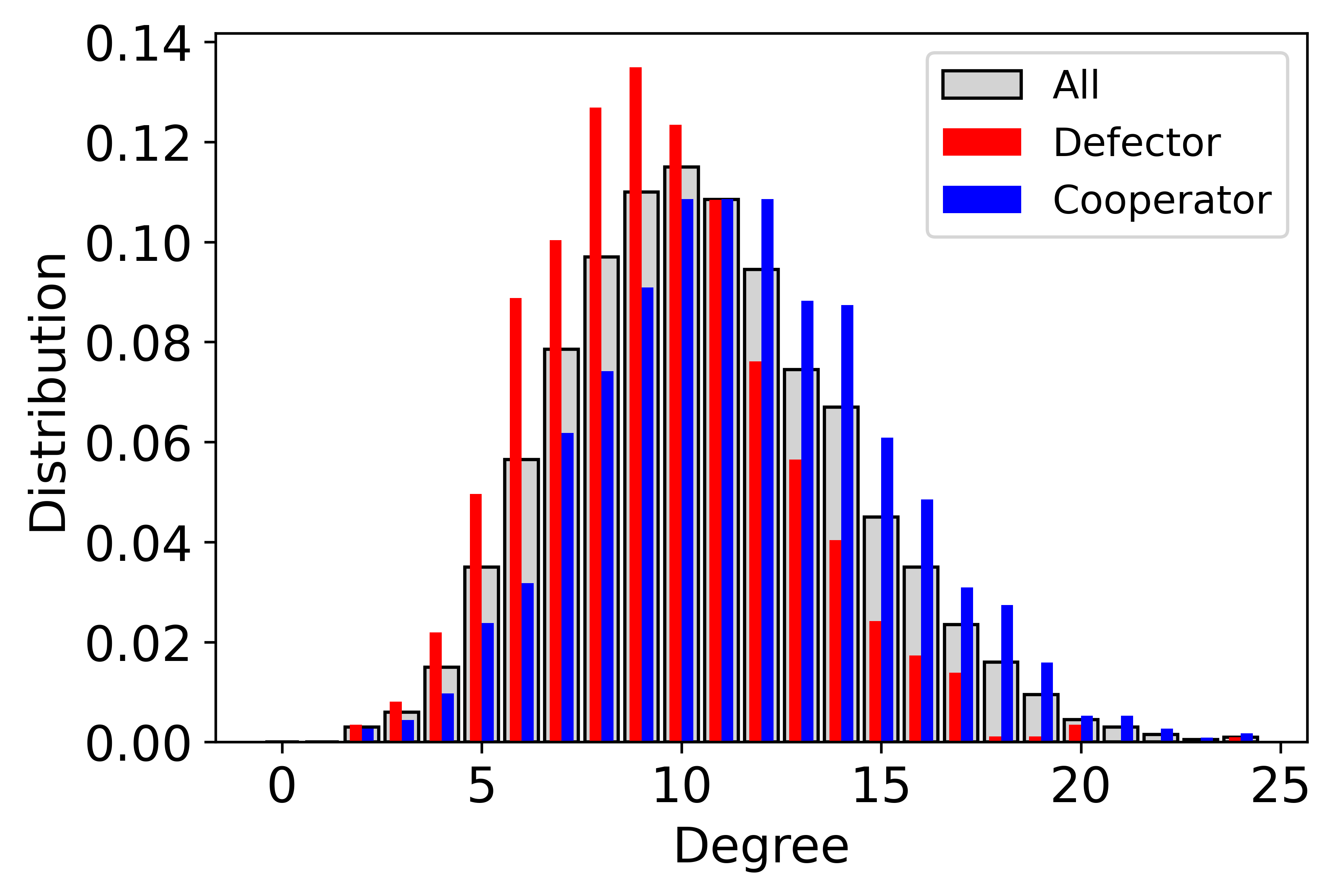}
    }
    \subfigure[]
    {
    \includegraphics[scale=0.46]{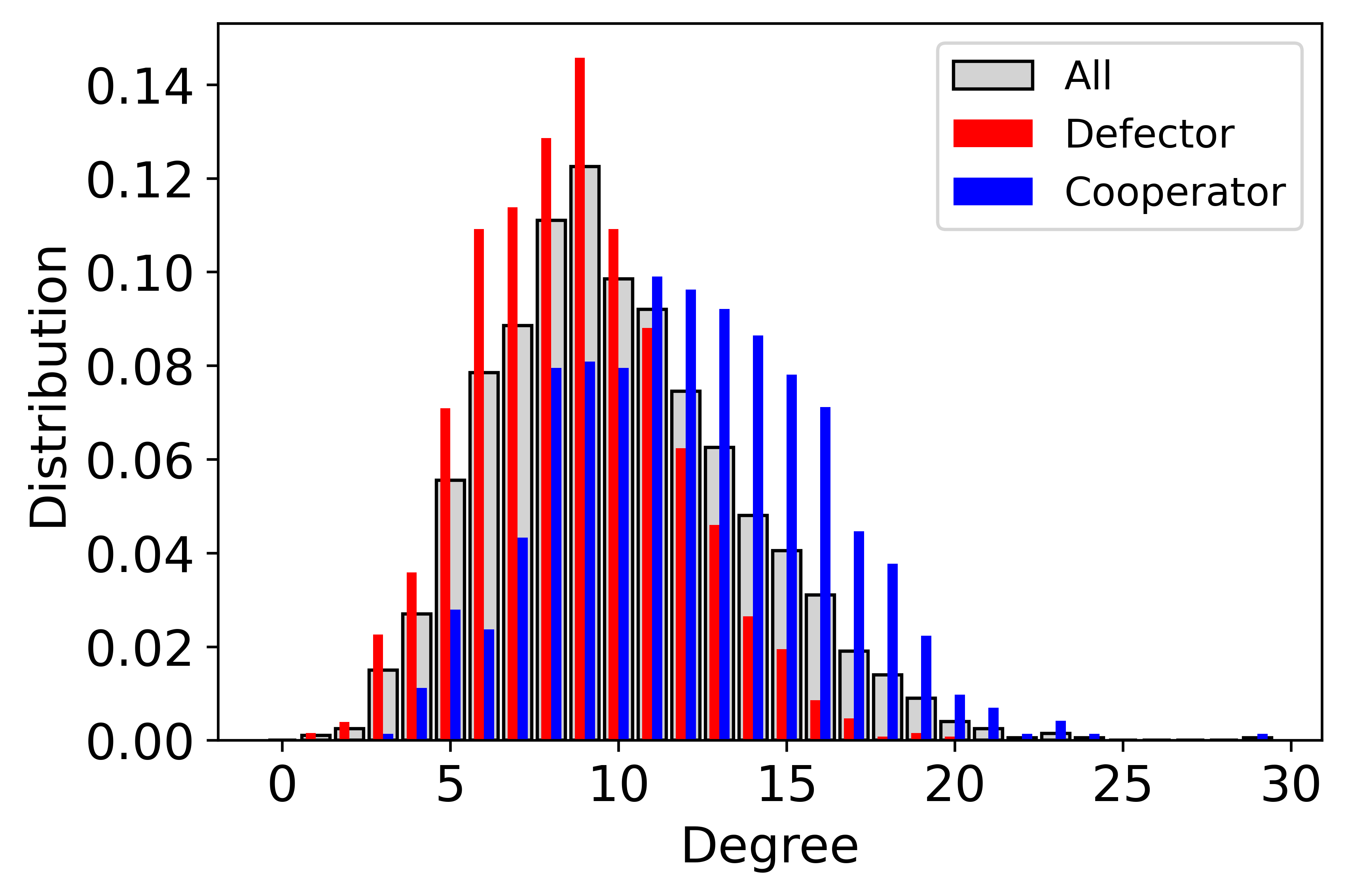}
    }
    \caption{\textbf{Degree distributions in stationary RG with varying average degrees of connectivity.} It displays the degree distributions for nodes in stationary state networks, differentiated by the average degrees of 4 and 10 within RG. Specifically, (a) presents a network with an initial average degree of $4$ and a cooperation rate of $0.2$, while (b) presents a network with an initial average degree of $10$ and the same cooperation rate. Moving to (c) and (d), the cooperation rate remains constant at $0.4$, but the initial average degree changes to $4$ and $10$, respectively. Similarly, (e) and (f) maintain a cooperation rate of $0.6$, with initial average degrees of $4$ and $10$, respectively. Each panel delineates the degree distributions for cooperators (blue) and defectors (red), along with all agents (gray).}
    \label{fig:5}
\end{figure*}

\begin{figure*}  
    \centering
    \hspace{-3.2mm}
    \subfigure[]
    {
    \includegraphics[scale=0.6]{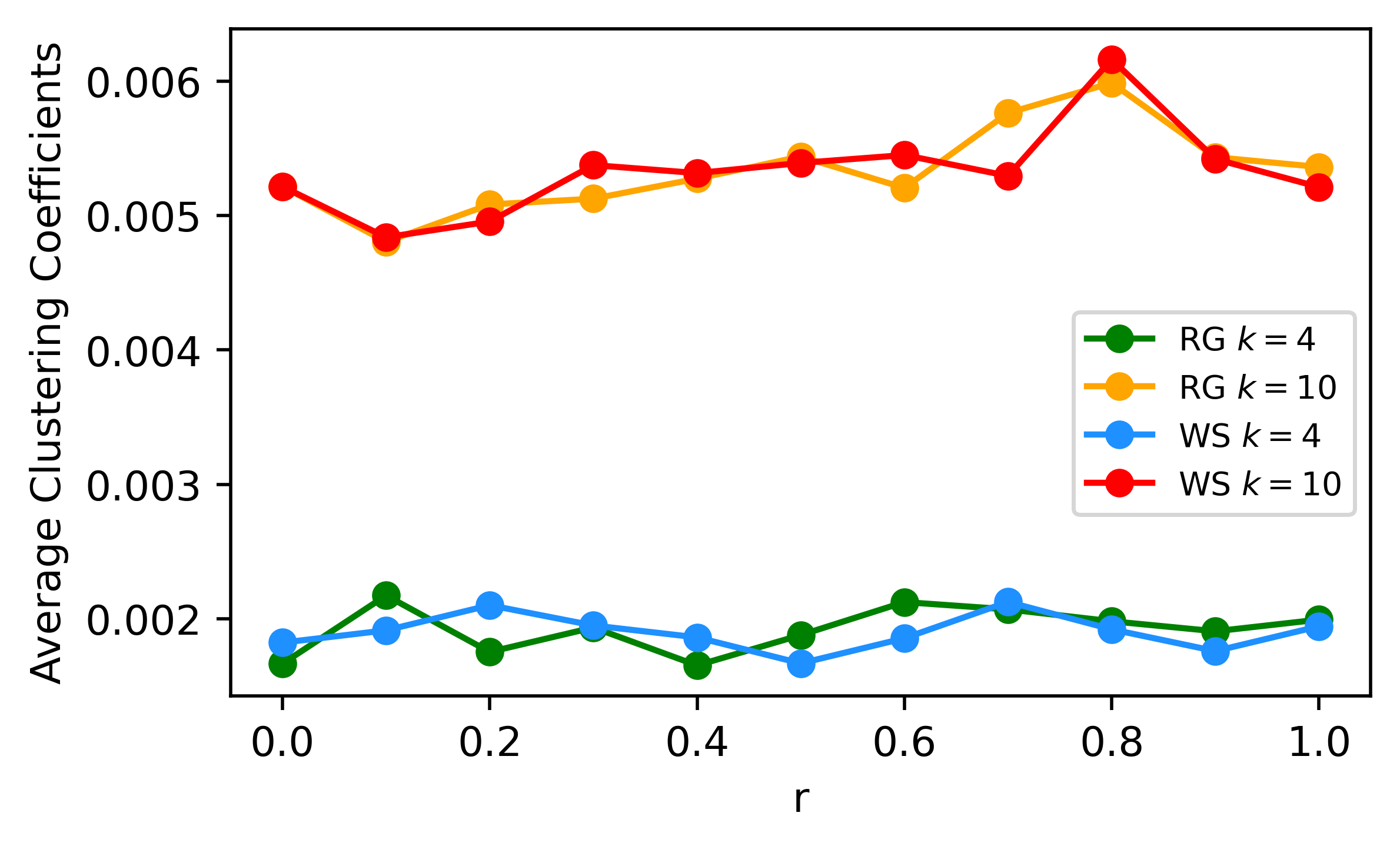}
    }
    \hspace{-0.61mm}
    \subfigure[]
    {
    \includegraphics[scale=0.6]{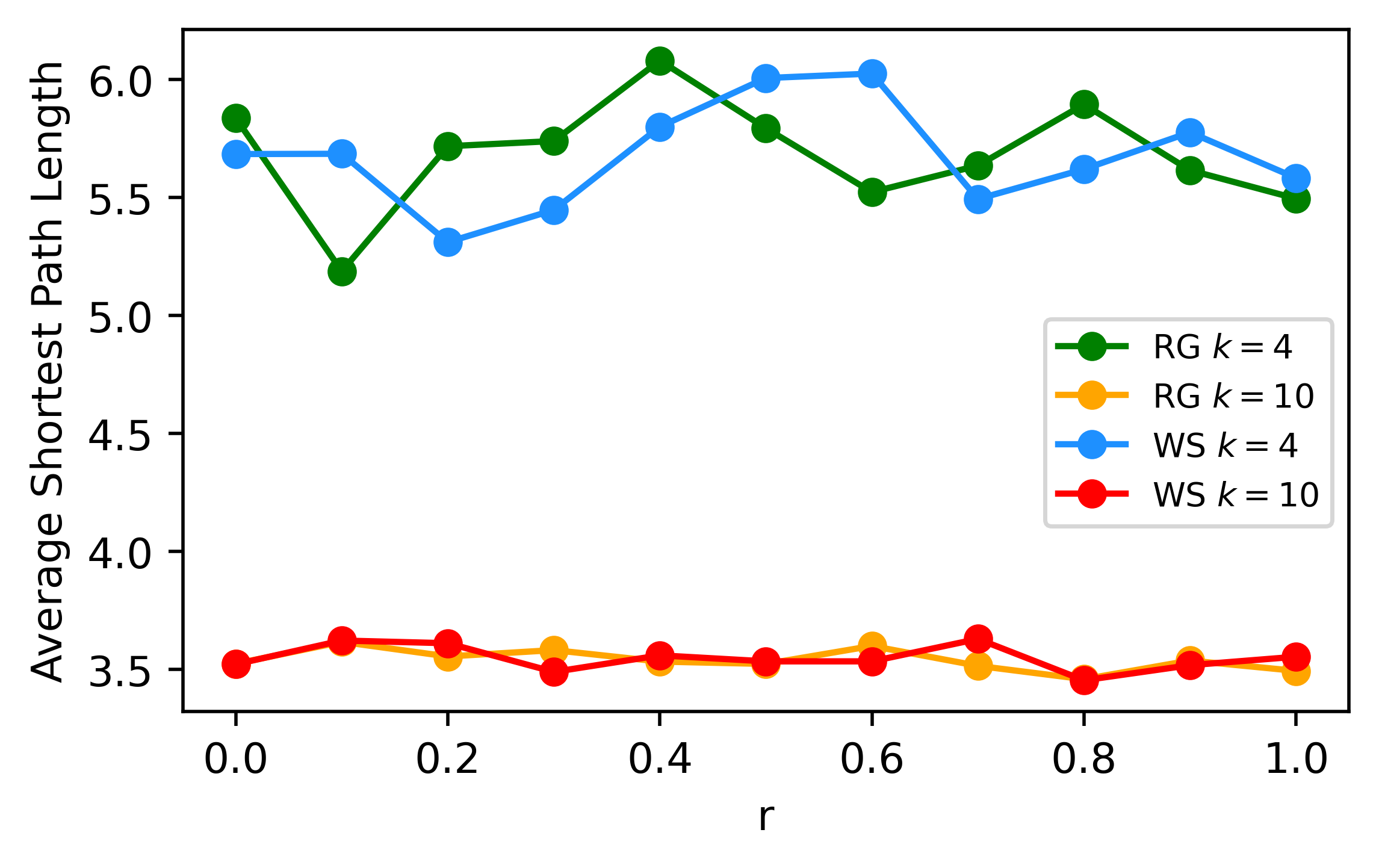}
    }
    \caption{\textbf{The average clustering coefficients and the average shortest path length of networks under different $r$.} Panel (a) displays the variation in average clustering coefficients for initial networks modeled as RG and WS  networks with mean degrees $k=4$ and $k=10$. Panel (b) illustrates the corresponding average shortest path lengths across these network models. It is evident from the results that the stationary state values of both the average clustering coefficients and the average shortest path lengths are predominantly influenced by the networks' average degree, while changes in the parameter $r$ do not significantly affect these metrics.}
    \label{fig:6}
\end{figure*}

To understand such an interaction, we investigate the topology of the resultant network among different edge types within adaptive networks over a spectrum of network configurations and $r$ values.  Firstly, our focus lies on comprehending the evolution of connection types, specifically cooperator-cooperator (C-C), cooperator-defector (C-D), and defector-defector (D-D) connections. As depicted in Fig.~\ref{fig:4}, we continue our investigation with RG and WS of average degrees 4 and 10, respectively, selecting $r$ values of 0.2, 0.4, and 0.6 to fully observe scenarios where cooperators dominate, defectors prevail, and the ratio of cooperators to defectors is approximate.

All results illustrate the proportions of the C-C, C-D, and D-D edge types after 300 iterations, averaging values over 10 independent runs, and ensuring that the network strategies have reached a stationary state. Across all parameters and network structures under scrutiny, we find that the dynamic proportions of these edge types reach a stationary state more rapidly than the evolution of strategies. In particular, a dominance of cooperative behaviors is evident when $r=0.2$, consistent with previous experimental findings indicating that the proportion of cooperators remains approximately 0.81 in various configurations. Therefore, at $r=0.2$, the initial network structure we explore here does not have a significant impact on the proportions of different edges. Specifically, there is a notable surge in the prevalence of C-C edges, which accounts for approximately 0.69 of all edges, while both D-D and C-D edges decline, with the proportion of D-D decreasing more rapidly to around 0.03 and the C-D diminishing to approximately 0.30.

In the case of RG and WS with $k=4$ and $r=0.4$, the fraction of cooperation is approximately 0.50, while it is approximately 0.57 for $k=10$. It is observable that in networks with a higher number of cooperators ($k=10$), the rate of increase in C-C edges and the rate of decrease in D-D edges are higher, and the disparity between stabilized C-C and D-D edges is more pronounced compared to the $k=4$ networks. Meanwhile, the C-D edges remain nearly constant at around 0.5 for $r=0.2$ in both network configurations. It suggests that the proportion changes of the three types of edges mainly depend on $f_c$ and are scarcely influenced by the initial network structures of RG or WS.

At $r=0.6$, the disparities between the RG and WS networks are minimal, but the gaps between networks with $k=4$ and $k=10$ are substantial, with stabilized cooperation proportions of 0.24 and 0.35, respectively. Due to the relatively higher proportion of cooperators in the $k=10$ network, C-D edges persist at 0.5, while the proportion of D-D edges surpasses that of C-C edges. Particularly for the sparser networks ($k=4$), C-D edges decrease to levels close to D-D edges, with a corresponding reduction in the proportion of C-C edges to around 0.09. The dynamic interplay of edge types under different conditions underscores the adaptive network's capability to reconfigure itself in response to evolving payoff landscapes, highlighting the importance of network topology in fostering or impeding cooperation.

Although we investigate the different types of edges between cooperators and defectors, the proportions of them are heavily influenced by the number of cooperators and defectors in the system. Consequently, we cannot directly infer the distribution patterns of cooperators and defectors based solely on these edge proportions. In addition, the series of experiments described above collectively indicate that various initial network types exert minimal influence on the evolutionary outcome of the adaptive final network, and this conclusion is further supported by our examination of degree distributions of simulations reaching stationary states in Fig.~\ref{fig:5}. Therefore, we present the degree distributions of all nodes in stationary state networks only for RG with $k$ values of 4 and 10.

In the range of parameters under investigation, regardless of whether $k$ equals 4 or 10, the degree distribution of all agents within the network exhibits a normal distribution. Evidently, cooperators possess a higher degree compared to defectors. When $r=0.2$, a condition characterized by a higher proportion of cooperators, the degree distributions between cooperators and defectors are relatively similar. However, with an increase in $r$ to 0.6, the disparity in degree distributions between cooperators and defectors becomes more pronounced, suggesting that in environments where defectors dominate, cooperators tend to form tighter clusters of cooperation to resist defector infiltration, while defectors tend to remain relatively isolated. The observation underscores that, under high values $r$, which indicate low cooperative tendencies, cooperators are particularly inclined to establish extensive partnerships. Furthermore, the influence of different initial network configurations on the eventually evolved network is negligible, indicating a strong self-organizing and adaptive capability of networks, which tend toward stable cooperative structures under diverse cooperation levels. The adaptability is particularly evident in the changes observed in the degree distributions of cooperators and defectors with varying $r$, compared to traditional static RG networks where all nodes maintain the same degree. It also reflects the strong adaptability of cooperators when facing numerous defectors, evident in its structural and connectivity patterns.

The network clustering coefficients and the shortest path length serve as crucial metrics to quantify the degree of clustering or the structure of the community present within the networks. To examine the variability of the average clustering coefficient and the average shortest path length in adaptive networks under stationary state conditions, we investigate the dynamics illustrated in Fig.~\ref{fig:6}. By systematically varying the parameter $r$, we emulate various levels of cooperation that shape node interactions and examine the influence of initial states on the clustering coefficients after the network reaches a stationary state. However, under certain parameter conditions, the network probably evolves such that some nodes become isolated, resulting in a disconnected network. In these cases, we calculate the two metrics mentioned above for the largest connected component of the network.

Interestingly, different values of $r$ result in different levels of cooperation in the stationary state. However, these variations do not exhibit a significant trend in altering the network's average clustering coefficient or average shortest path length. Fig.~\ref{fig:6} (a) illustrates that the average degree of the four previously studied network structures plays a decisive role in the stationary state outcomes of the average clustering coefficient of the network, while the differences in the initial network structures have minimal impact. Despite the initial average degrees of 4 and 10 in RG have average clustering coefficients of 0.0008 and 0.004, respectively, and in WS that have average clustering coefficients of 0.3713 and 0.4871, all network structures, except RG with $k=10$, evolve toward smaller clustering coefficients. 

Fig.~\ref{fig:6} (b) shows that regardless of the value of $r$ for both $k=4$ and $k=10$ in WS and RG, the evolution of the system has minimal impact on the average shortest path length of the network and lacks a discernible pattern. Initially, RG with average degrees of 4 and 10 has average shortest path lengths of 6.26 and 3.59, respectively, while WS has average shortest path lengths of 9.85 and 4.94. Across the network structures explored in this study, the networks generally evolve toward smaller average shortest path lengths, indicating that the networks become more efficient in terms of communication and connectivity as they evolve, despite the initial structural differences.

In summary, networks with an average degree of 10 exhibits significantly higher clustering coefficients and shorter path lengths compared to those with an average degree of 4. It can be explained by the fact that a higher average degree provides more connections per node, which enhances the likelihood of forming tightly-knit clusters and reduces the number of steps required to traverse the network. Thus, networks with higher average degrees tend to become more clustered and efficient over time, reflecting the dynamic interplay between degree distribution and network topology.

\section{Conclusions and future research}
\label{sec:conclusion}

In this study, we propose and investigate an adaptive network based on the snowdrift game, aimed at gaining deeper insights into the evolution of cooperative behavior in dynamic networks driven by the maximization of payoff by agents and its influence on the surrounding neighborhood structure and the entire network. Building upon this, we introduce the concept of the agent's disassociation tendency toward its neighbors based on payoffs, defining the probability of edges between neighbors no longer being connected, while stipulating that agents whose edges are cut can randomly select a replacement non-neighbor agent.

Through experimental simulations, we analyze the effects of different initial network configurations (RG and WS with average degrees of 4 and 10), as well as the cost-benefit ratio ($r$), on the evolution of cooperative behavior, revealing the impact of dynamic adaptability of the network on cooperative behavior. Initially, we observe that compared to traditional static networks, adaptive networks are more likely to reach states of pure cooperation or pure defection at extreme values of $r$. We find that networks with higher average degrees ($k=10$) exhibit relatively higher levels of cooperation compared to those with lower degrees ($k=4$). Additionally, both of them promote cooperative behavior across a wider range of $r$ values compared to statically configured networks of the same type. Understanding the structural distribution of cooperators and defectors within networks is crucial to comprehending the dynamics of cooperation and defection in social systems. We also investigate the intricate interplay between cooperative and non-cooperative behaviors through an analysis of network structures. 

A detailed analysis is conducted on the types of network connections, specifically C-C, C-D, and D-D connections in adaptive networks. We show that the evolution of edges reaches a stationary state more rapidly than the evolution of strategies. Moreover, we can infer that the distribution ratio of edges almost directly determines the final proportion of cooperators in the network. In addition, the distributions of these connection types were analyzed in different network configurations and $r$ values. In systems with a higher prevalence of defectors, cooperators demonstrate a marked propensity to establish extensive partnerships to counteract the incursion of defectors. Interestingly, the average clustering coefficient and the shortest path length of the network do not exhibit a clear pattern of variation with changes in cooperation levels. Instead, it is primarily influenced by the initial structure of the network.

Further research can explore the implications of more diverse network configurations and introduce modifications to adaptive rules to better reflect real-world complexity, as done in previous evolutionary games fed with real data~\cite{Chica21CIM}. Additionally, in terms of strategy updating, we assume that fully rational individuals will only imitate the strategies of neighbors who achieve higher payoffs. However, there are other rules for updating the strategy that warrant a thorough investigation~\cite{Chica19CNSNS}. Finally, the current model posits that agents select new neighbors randomly. Exploring more complex rules for neighbor selection could enrich the model's applicability to understanding human social behaviors. The potential to apply this adaptive network model to other types of games and social settings, such as the generalized pairwise game or asymmetric games~\cite{mcavoy2015asymmetric}, opens new avenues for exploring the intricate dynamics of cooperation within complex networks.

\section*{Acknowledgments}
X. Xiong, Y. Yao and M. Feng are supported by grant No. 62206230 funded by the National Natural Science Foundation of China (NSFC), and grant No. CSTB2023NSCQ-MSX0064 funded by the Natural Science Foundation of Chongqing. M. Chica is supported by grant EMERGIA21\_00139 funded by Consejería de Universidad, Investigación e Innovación of the Andalusian Government as well as CONFIA (PID2021-122916NB-I00, MCIN/AEI/10.13039/501100011033 and FEDER ``Una manera de hacer Europa'').


\end{document}